# Terahertz-to-infrared converters for imaging the human skin cancer: challenges and feasibility

Kamil Moldosanov,[a,*] Alexander Bykov,[b] Nurlanbek Kairyev,[a] Mikhail Khodzitsky,[c,*] Grigory Kropotov,[c] Valery Lelevkin,[a] Igor Meglinski,[b,d,*] Andrei Postnikov,[e,*] and Alexey Shakhmin[c]

[a]Kyrgyz-Russian Slavic University, Bishkek, Kyrgyzstan
[b]University of Oulu, Oulu, Finland
[c]Tydex LLC, Saint Petersburg, Russia
[d]College of Engineering and Physical Sciences, Aston University, Birmingham, UK
[e]Université de Lorraine, LCP-A2MC, Metz, France

**Abstract**

**Purpose:** The terahertz (THz) medical imaging is a promising noninvasive technique for monitoring the skin's conditions, early detection of the human skin cancer, and recovery from burns and wounds. It can be applied for visualization of healing process directly through clinical dressings and restorative ointments, minimizing the frequency of dressing changes. The THz imaging technique is cost effective, as compared to the magnetic resonance method. Our aim was to develop an approach capable of providing better image resolution than the commercially available THz imaging cameras.

**Approach:** The terahertz-to-infrared (THz-to-IR) converters can visualize the human skin cancer by converting the latter's specific contrast patterns recognizable in THz radiation range into IR patterns, detectable by a standard IR imaging camera. At the core of suggested THz-to-IR converters are flat matrices transparent both in the THz range to be visualized and in the operating range of the IR camera; these matrices contain embedded metal nanoparticles, which, when irradiated with THz rays, convert the energy of THz photons into heat and become nanosources of IR radiation detectable by an IR camera.

**Results:** The ways of creating the simplest converter, as well as a more complex converter with wider capabilities, are considered. The first converter is a gelatin matrix with gold 8.5 nm diameter nanoparticles, the second is a polystyrene matrix with 2 nm diameter nanoparticles from copper–nickel MONEL® alloy 404.

**Conclusions:** An approach with a THz-to-IR converter equipped with an IR camera is promising in that it could provide a better image of oncological pathology than the commercially available THz imaging cameras do.

## 1 Introduction

On the scale of electromagnetic (EM) waves, there is a frequency band interesting for potential application in cancer cell imaging. This is the so-called terahertz (THz) range which spans the gap between radio waves and infrared (IR) radiation. THz radiation has properties inherent in neighboring bands: like radio waves, it penetrates many optically opaque media, and, like IR rays, can be refracted and focused by lenses.

For a number of years, we developed and theoretically evaluated the parameters of the so-called terahertz-to-infrared (THz-to-IR) converters [1-9] (see Table 1). Should the production of these devices be established, they would allow the conversion of invisible THz radiation into IR rays that could be visualized with commercial IR cameras. This would allow them to be used in devices for imaging human skin cancer.

The THz-to-IR converters in question consist of matrices, transparent both in the range of THz radiation to be visualized and in the operating range of the IR camera, the matrices being "stuffed" with metal nanoparticles (NPs) (see Fig. 1). These latter, when irradiated with THz rays, convert the energy of THz photons into heat and become sources of IR radiation, which can be further on detected by an IR camera.

Table 1. Estimated characteristics of the THz-to-IR converters to be used with Mirage 640 P-Series IR camera. All data are given for the practically significant value of the emissivity factor $\alpha = 0.5$.

| Ref. | Parameter | | | | | |
|---|---|---|---|---|---|---|
| | NP material | NP diameter (nm) | Matrix material | NP heating / cooling times (ns) | Power required to heat the NP to be seen by the IR camera (nW) | Frequency of operating radiation (THz) |
| [1, 2] | Ni | 2.4 | Gelatin | 13/13 | 0.13 | 0.35–0.55 |
| [3] | Cu-Ni alloy | 2.5 | | 72/78 | 0.13 | |
| [4] | Au | 2.4 | Teflon® | 79/71 | 0.11 | 0.38; 4.2; 8.4 |
| [5–7] | | 8.5 | | 300/1281 | 0.34 | |
| [5–7] | | 8.5 | Silicon | 0.17/1.16 | 98.4 | |
| [8] | | 6.0 | Teflon® | 230/1190 | 0.28 | 0.7 |
| | | 9.5 | | 330/1360 | 0.44 | 0.41 |
| | | 14.65 | | 430/1390 | 0.68 | 0.24 |
| | | 25.4 | | 560/1580 | 1.2 | 0.14 |
| [9] | | 8.5 | Polyvinyl chloride | 306/1314 | 0.21 | 0.38; 4.2; 8.4 |
| This article | | | Gelatin | 407/1718 | 0.27 | 0.35–0.55 |
| | | | Polystyrene | 319/1339 | 0.18 | 0.38; 4.2; 8.4 |
| | | | Polypropylene | 292/1255 | 0.25 | |
| | 50at.%Au50at.%Pd alloy | 16 | Gelatin | 607/2102 | 0.51 | 0.35–0.55 |
| | MONEL alloy 404 | 2 | Gelatin | 83/942 | 0.06 | 0.35–0.55 |
| | | | Teflon® | 61/690 | 0.08 | 0.4; 5; 10 |
| | | | Polystyrene | 65/734 | 0.04 | |
| | | | Polypropylene | 59/675 | 0.06 | |
| | | | Polyvinyl chloride | 62/707 | 0.05 | |

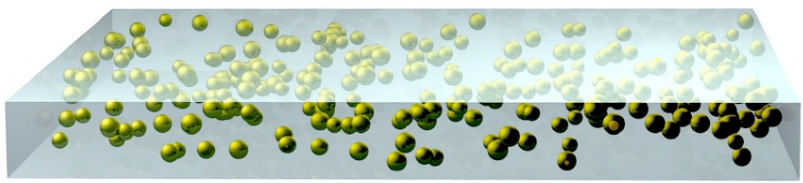

Fig. 1. The "working element" of a THz-to-IR converter: a matrix with embedded metal NPs.

In this work, we focus on the use of THz-to-IR converters for imaging the human skin cancer. Two circumstances, the fact that the water content of cancer cells is higher than that of normal cells [10,11], and the fact that the reflectivity of the THz radiation (and hence the image contrast of the cancer area) increases as the temperature of the water in cancer cells increases [12-16], favor the development of the reflection geometry approach to the THz medical imaging of a cancerous biological tissue (for in vivo studies, Fig. 2).

To heat water in cancer cells, the gold nanoparticles (GNPs), like targeted agents in photothermal therapy, are premediatedly delivered into the cancer cells, but not into the normal cells. The approach relies on the fact that the targeted agents, antibody-conjugated GNPs, are accumulated in cancerous tissue more efficiently than in the normal one. Thereupon the tumor is non-invasively treated by irradiating with near-infrared (NIR) laser beam at ~650–1350 *nm* wavelength; this is the so-called "therapeutic window" where light has its maximum depth of penetration into the tissues. Under irradiation, the surface plasmons are excited in the GNPs; as the plasmons decay, the water around the GNPs in the cancer cells heats up. In consequence, the cancer cells start to reflect the incident THz radiation even more efficiently [12-16], and thus can be more readily visualized by a highly sensitive IR camera (Fig. 2).

The feasibility of the THz imaging of the body with skin cancer in reflection geometry has been demonstrated in works by Woodward et al. [17,18]. The pioneering studies [19-21] revealed the possibility of THz imaging of basal cell carcinoma.

Wallace et al. [21] and Pickwell and Wallace [22] showed that when biological tissue is irradiated with THz radiation, the maximum difference in the refractive index between the diseased and normal tissue occurs in the frequency band 0.35–0.55 THz (wavelengths, respectively: 857–545 μm). In the same band, there is the maximum difference in the absorption of diseased and normal tissue, which provides the best image contrast and helps to identify tumor margins. We briefly reviewed articles relevant to our problem [23-27] (see Table 2) and current

Table 2. Summary of findings from some THz bioeffects studies relevant to our problem

| Ref. | What is noteworthy in the study | Irradiance (mW·cm$^{-2}$) | Frequency (THz) | Exposure durations | Findings |
|---|---|---|---|---|---|
| [23] | (i) A frequency used was 0.38 THz, which we chose for heating GNPs (see Subsecs. 3.1.1 and 3.1.2); (ii) human skin cells were studied. | 0.03–0.9 | 0.38 | 2 and 8 hours | The 0.38 THz radiation does not induce genomic damage. |
| [24] | (i) A frequency range was used containing the frequency of 0.38 THz chosen for heating GNPs; (ii) human skin cells were studied. | $\leq 5 \cdot 10^{-3}$ | 0.3–0.6 | 94 hours | No cellular damage or changes to cell proliferation were observed. |
| [25] | (i) Frequencies sweeping from 0.07 to 0.3 THz were used, i.e. with upper frequency near to 0.38 THz, chosen for heating GNPs; (ii) human skin fibroblast cells were studied. | $7 \cdot 10^{-5} - 3.8 \cdot 10^{-4}$ | 0.07–0.3 | 3, 70, and 94 hours | No changes in cell activity and no cytotoxicity were found. |
| [26] | (i) A current safety limit of incident power density for local exposure, 2 mW·cm$^{-2}$ [28], was used; (ii) human skin cells were studied. | 2 | 0.106 | 2, 8, and 24 hours | No changes in proliferation rate and no abnormalities in the genetic apparatus. |
| [27] | (i) Exceeding a current safety limit of incident power density for local exposure, 2 mW·cm$^{-2}$ [28], for general public, by 1.5 times; (ii) albino rats were researched. | 3 | 0.15 | 0.25, 0.5, and 1 hour | Rats exposed to radiation for 1 hour had increased levels of depression and enhanced platelet aggregation. Rats exposed for less than 1 hour, did not exhibit either effect. |

ICNIRP safety limits [28-30] and concluded that one should respect the following conditions of skin tussue exposure to THz radiation: irradiance $\leq$ 2 mW·cm$^{-2}$; exposure durations $\leq$ 10 s; frequency 0.38 THz, providing contrast margin of the cancerous tissue (see Subsecs. 3.1.1 and 3.1.2).

In the setup in Fig. 2, the THz rays reflected from the tumor are focused by a lens on the THz-to-IR converter's matrix, creating in this plane a THz image of the tumor. Further on, the metal NPs within the THz-to-IR converter absorb the THz radiation, generate heat and become IR sources, producing an image detectable by the IR camera.

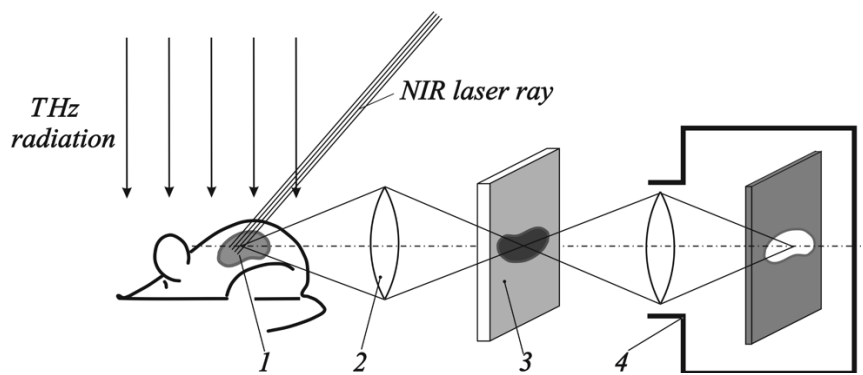

Fig. 2. The reflection geometry for *in vivo* imaging of oncopathology using a near-IR laser to excite surface plasmons in GNPs in a tumor (for heating water in cancer cells); 1 – tumor, 2 – THz objective with a magnification $M_1$, 3 – THz-to-IR converter, 4 – highly sensitive IR camera with magnification $M_2$. See [4] for details.

## 2 Selection of IR cameras, lens system and combinations of materials to be used in THz-to-IR converters

A natural question may arise, why do we propose to use not a commercially available THz imaging camera, but a THz-to-IR converter with an IR camera to visualize oncopathology?

A comparison of the pixel pitch sizes of the Mirage 640 P-Series IR thermal imaging camera [31] (hereinafter referred to as Mirage; it is the most sensitive procurable IR imaging camera, with the temperature sensitivity of 12 mK at 300 K) and commercially available THz imaging cameras (see Table 3) underlines the advantages of our approach in the perspective of imaging of oncopathology with acceptable resolution.
The THz imaging RIGI Camera has the same pixel pitch size as the Mirage, and its pixel array is even larger; however, its operating frequency by far exceeds the frequencies suitable for biomedical research, 0.35–0.55 THz. Other THz imaging cameras have larger pixel pitch sizes than the Mirage camera, and their pixel arrays are smaller. Therefore, they would not provide as good resolution as the THz-to-IR converter equipped with the Mirage camera promises.

In Ref. 32, Table 1 shows that the sensitivities per pixel of the IR/V-T0831 and TZcam cameras are no worse than 0.1 and 0.02 nW, respectively. Our calculations of the powers required to heat the 2 nm diameter NP of the copper-nickel MONEL 404 alloy (hereinafter referred to as monel) in plastic and gelatin matrices, so that it can be recorded by an IR camera after heating, gave values of 0.04–0.08 nW (see Table 1). Assuming that we have at least one NP located in the volume projected onto a size of a single detector pixel (see Ref. 5 and Subsec. 5.2 for details), the pixel sensitivity of the THz-to-IR converter based on 2 nm diameter monel NPs is of the same order of magnitude as the pixel sensitivity of THz imaging cameras. Thus, our approach using the THz-to-IR converter and the IR camera will hold in what concerns sensitivity.

Table 3. Comparison of parameters of the Mirage 640 P-Series IR camera and commercially available THz imaging cameras

| Imaging camera | Manufacturer | Pixel pitch size | Pixel array | Operating frequency (THz) | Ref. |
|---|---|---|---|---|---|
| Mirage 640 P-Series | Infrared Cameras Inc. | 15 μm | 640×512 | 60–200 | [31] |
| RIGI Camera | Swiss Terahertz | 15 μm | 1920×1080 | 4.6 | [32,33] |
| IR/V-T0831 | NEC | 23.5 μm | 320×240 | 1–7 | [32,34] |
| MICROXCAM-384i-THz | INO | 35 μm | 384×288 | 0.094–4.25 | [32,34,35] |
| IRXCAM-THz-384 | INO | 35 μm | 384×288 | 0.1–7.5 | [36] |
| TZcam | i2S | 50 μm | 320×240 | 0.3–5 | [32,34,37] |
| THXCAM-160 | INO | 52 μm | 160 x 120 | 0.69; 2.52 | [38] |
| IRXCAM-160 THz | INO | 52 μm | 160 x 120 | 2.54 | [39] |
| Pyrocam IV | Ophir Photonics | 75 μm | 320×320 | 0.1–283 | [34,40] |
| OpenView | Nethis | 80 μm | 640×512 | 0.1–3000 | [34,41] |
| OpenView | Nethis | 170 μm | 320×256 | 0.1–3000 | [34,41] |
| TERACAM | Alphanov | 1 mm | 40×40 | 0.1–30 | [32] |
| Tera-4096 | TeraSense | 1.5 mm | 64×64 | 0.05–0.7 | [42] |
| Tera-1024 | TeraSense | 1.5 mm | 32×32 | 0.05–0.7 | [32,34,42] |

In initial works, we suggested to use NPs of nickel [1,2] and Cu-Ni alloys [3], which have high density of electronic states near the Fermi level, hinting for sufficiently high absorption capacity. Then, it became clear that it would be more convenient to use commercially available GNPs, which are produced with a fairly small spread of sizes around the nominal one.

We have considered different combinations of materials to be used in THz-to-IR converters. Table 1 specifies the parameters of some of them, chosen so, in order to be specific, as to make use of the temperature sensitivity of ≈12 mK, inherent for the Mirage camera [31]. The numerical estimates demonstrate that the converters so suggested might operate on a real-time mode, since the heating and cooling times of the selected NPs are sufficiently short.

The purpose of this review about THz-to-IR converters is not only to revive the theoretical studies we have earlier presented on the choice of matrix and NPs materials for them, with the estimation of corresponding parameters, but also to introduce new approaches, with an ambition to simplify the converter manufacturing process. Unfortunately, so far we have not seen an experimental verification of our idea, not least because of the technological difficulties of introducing NPs into the matrix. Therefore, as an ultimate simplification of the approach, we suggest to choose gelatin as the matrix material, along with commercially available GNPs. The latter are supplied as a sol, a colloid made out of GNPs in a water medium, and gelatin is a water soluble protein. Then, we hope, the procedure for introducing GNPs into a matrix would simply be reduced to draining two liquids and sonicating them in order to achieve a uniform distribution of GNPs in the matrix.

In this study, we have compared two THz-to-IR converters in the form of gelatin matrices containing the 8.5 nm diameter GNPs or the 2 nm diameter monel NPs. We considered two physical mechanisms for the conversion of THz radiation into heat at frequencies favorable for biomedical research: in 8.5 nm diameter GNPs, mainly due to the excitation of transverse and longitudinal phonons, and in the 2 nm diameter monel NPs, mainly due to electron-electron scattering.

Why a 8.5 nm diameter is suggested for GNPs? There are four reasons for this:

(1) There will be no background cooling of the NP due to spontaneous emission of THz photons, which, according to our hypothesis (see Appendix A), may occur at GNPs diameters less than ≈8 nm.

(2) Using integers $m_{el}$ = 4 and $n_{vm}$ = 1 for a medically favorable frequency $v$ = 0.38 THz, formula (5) (see Subsec. 3.1.2) gives an acceptable GNP diameter $D \approx$ 8.5 nm.

(3) GNPs of this diameter can be heated by absorbing EM radiation of both 0.38 THz frequency, favorable for biomedical research, and radiation of increased THz frequencies (4.2 and 8.4 THz), providing better resolution of the image of oncopathology (see Table 4). The heating occurs as excitation of both longitudinal and transverse phonons (see Sec. 3). A comparison of the distribution graphs of longitudinal phonons in bulk gold and in a GNP showed (see Fig. 2 in Ref. 43) that they are similar, and the peak of the distribution is at ≈4.2 THz. That is, there are a lot of phonons with a frequency of 4.2 THz in a GNP, and 8.4 THz is the twice the frequency of the distribution peak. Therefore, one 8.4 THz photon may excite two dominant phonons with a frequency of 4.2 THz, since there are no longitudinal phonons with a frequency of 8.4 THz in a GNP. The choice of operating frequencies of 4.2 and 8.4 THz would increase the efficiency of heating the GNP. Therefore, the same THz-to-IR converter could be used to map the same fragment of a human skin: at the frequency 0.38 THz with enhanced contrast between cancerous and normal tissues, and at higher frequencies with enhanced resolution. A comparison of corresponding images would help to better diagnose the pathology.

In this approach, we rely on the following fact: according to Fig. 8 in Ref. 1, under normal conditions the transmittances of the 1.2-m air path at wavelengths of 35.7 μm (8.4 THz) and 71.3 μm (4.2 THz) are ~40–85% and ~30–50%, respectively. If the air path is reduced (by bringing the 4.2 and 8.4 THz sources closer to human skin tissue), then the air transmittances will be higher, and visualization at these frequencies will become feasible.

(4) Finally, such GNPs are mass-produced, which would simplify the technology of manufacturing a THz-to-IR converter with a gelatin matrix.

The reasons to use, in case of monel, the NPs of much smaller 2 nm diameter are the following.

(i) In the electron density of states of copper-nickel alloys of ~50at.%Cu50at.%Ni concentration, there is a pronounced peak at the Fermi level, persisting even if the local variations of interatomic distances ought to be strongest at these intermediate concentrations.

This causes an increase in the intensity of electron scattering (a decrease in the mean free path $l_{mfp}$ of electrons), which is favourable for efficiently converting the energy of THz photons into heat. In the monel alloy (47wt%Cu53wt%Ni), the minimum $l_{mfp} \approx$ 0.85 nm is achieved.

(ii) Due to low enough $l_{mfp}$ in the monel alloy, the NP diameter $D$ can be chosen small (≈2 nm), which reduces the NP volume heated by THz photons and, consequently, increases the sensitivity of the THz-to-IR converter in terms of detectable THz power.

(iii) The rule-of-thumb condition for the electron energy to be released inside the 2 nm diameter monel NP volume reads: $l_{mfp} \leq (D/2)$, in the assumption that the excitation of a Fermi electron by a THz photon happens at the NP's center.

(iv) The monel alloy offers an optimal relation between the electron DOS at the Fermi level and $l_{mfp}$. This circumstance, along with arguments (ii) and (iii), works for an efficient conversion of THz radiation into IR radiation.

(v) By force of the condition $l_{mfp} \leq (D/2)$, a monel NP would absorb in the entire THz range (0.1–10 THz), hence throughout all the frequencies of practical importance mentioned above.

(vi) The THz-to-IR converter based on 2 nm monel NPs, coupled with an IR camera, ought to have the same sensitivity per pixel of the IR camera detector as commercially available THz imaging cameras do, but the resolution promises to be better than that of those.

The choice of matrix material depends on the type of IR camera to be used to capture images produced by the THz-to-IR converter in IR rays. The matrix material should be transparent both in the range of THz radiation to be visualized and in the operating range of the IR camera. In this regard, one should look at IR cameras operating in the wavelength range containing the wavelength of 10 μm, which corresponds to the peak of the emissivity distribution of a heated NP at a temperature of 300 K. In Table 4, the data of commercially available IR cameras, operating within the range near the 10 μm wavelength, are given (here $\Delta T$ is the temperature sensitivity of the IR camera).

Table 4. Parameters of commercially available FLIR IR cameras manufactured by Teledyne FLIR.

| Imaging IR camera | Pixel pitch size (μm) | Pixel array | Operating wavelengths (μm) | $\Delta T$ (mK) | Ref. |
|---|---|---|---|---|---|
| T860 | 12 | 640×480 | 7.5–14.0 | 40 | [44] |
| A655sc | 17 | | | 30 | [45] |
| T650sc | 17 | | | 20 | [46] |

The T860 camera's detector has a small pixel size of 12 μm, which promises higher image resolution, but its temperature sensitivity is not so high (40 mK). Detectors of the A655sc and T650sc cameras have a larger pixel size, 17 μm, while the T650sc camera has a higher temperature sensitivity, 20 mK.

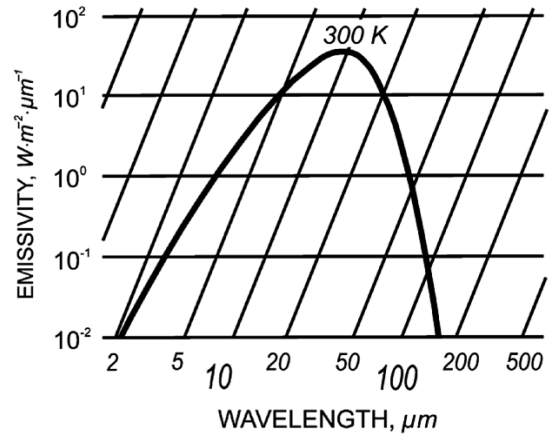

Fig. 3. The emissivity of a heated black body at temperature of 300 K.

According to Fig. 3, within the wavelengths corresponding to the operating ranges of the Mirage and FLIR IR cameras, the emissivity of a heated NP at temperature of ≈300 K is significantly different. It is not yet obvious which of the cameras and, accordingly, which matrix material would provide a better image of oncopathology.

Mirage [31] and FLIR A6700 MWIR [47] IR cameras can be used in combination with all matrix materials considered in Table 1. However, FLIR IR cameras [44-46], listed in Table 4, can work only in combination with polypropylene and Teflon® films (note that, for these latter, the heating/cooling times and the power required to heat the NPs will differ from the values shown in Table 1). Polystyrene, gelatin, and polyvinyl chloride films cannot be used with the mentioned FLIR IR cameras [44-46]: they are not transparent in the ~10 μm wavelength region, the operating range of FLIR cameras (see, respectively, Figs. 5 and 9, and Fig. 2 in Ref. 9).

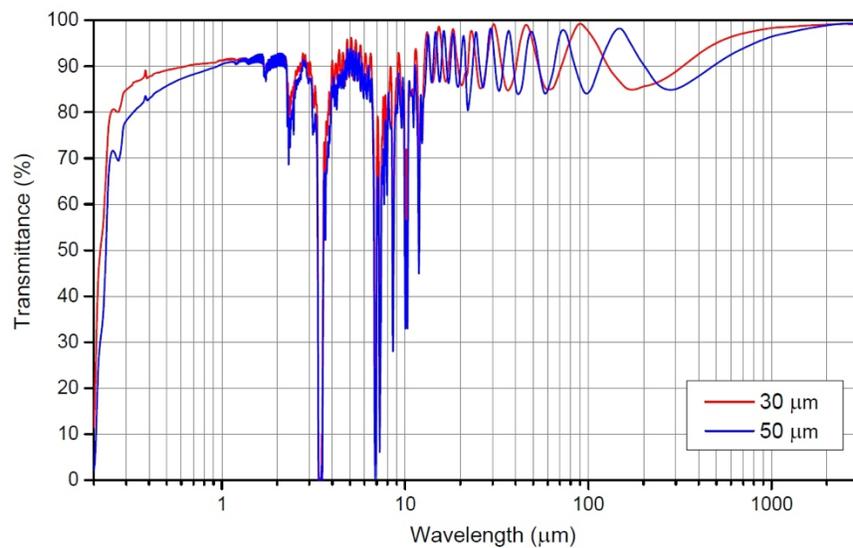

Fig. 4. Transmission spectrum of polypropylene films 30 μm (red) and 50 μm (blue) thick.

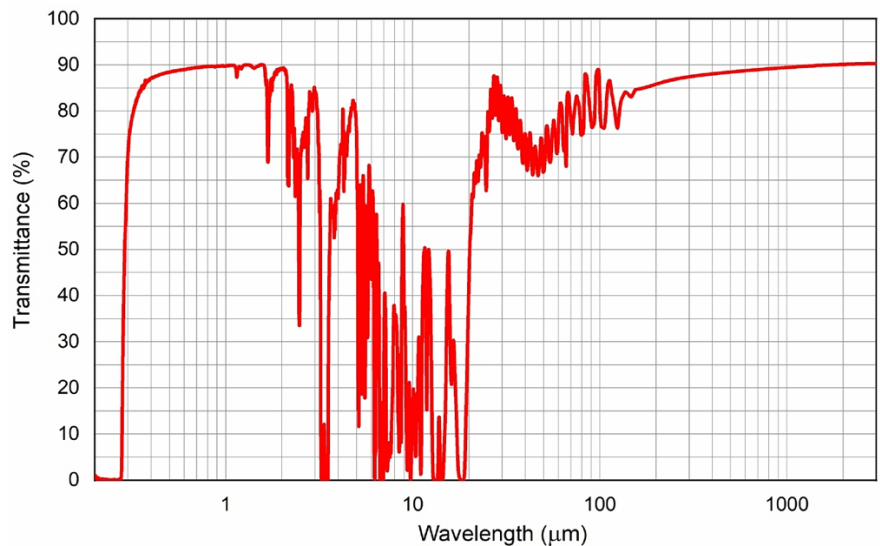

Fig. 5. Transmission spectrum of a polystyrene film 220 μm thick.

Table 5. Comparison of estimated characteristics of the THz-to-IR converters to be used with FLIR IR cameras from Table 4. All data are given for the practically significant value of the emissivity factor $\alpha = 0.5$.

| NP material | NP diameter (nm) | Matrix material | NP heating / cooling times (ns) | | Power required to heat the NP to be seen by the IR camera (nW) | |
|---|---|---|---|---|---|---|
| | | | $\Delta T = 20$ mK | $\Delta T = 30$ mK | $\Delta T = 20$ mK | $\Delta T = 30$ mK |
| Au | 8.5 | Teflon® | 300/1305 | 300/1309 | 0.56 | 0.84 |
| | | Polystyrene | 319/1389 | 319/1393 | 0.30 | 0.45 |
| | | Polypropylene | 292/1272 | 292/1275 | 0.41 | 0.61 |
| MONEL alloy 404 | 2 | Teflon® | 61/691 | 61/691 | 0.13 | 0.195 |
| | | Polystyrene | 65/735 | 65/735.5 | 0.07 | 0.105 |
| | | Polypropylene | 59.5/674 | 59.5/674 | 0.095 | 0.14 |

The following considerations help to specify the lens system and sizes of the THz-to-IR converter.

The objective of the IR camera ensures the close-up operation mode with magnification $M_2 = 1$. For IR cameras, which are kept in mind [31,47], this imposes a short distance (≈22–23 mm) between the "object" (i.e., the THz-to-IR converter's matrix with GNPs) and the objective's edge, as well as small sizes (9.6 mm × 7.7 mm) of the "object" (9.6 mm = 640 pixels × 15 μm; 7.7 mm ≈ 512 pixels × 15 μm). The THz-to-IR converter being so small seems technologically advantageous. At small sizes of the converter's matrix, the distortions caused by the first lens (the THz objective) would be small, too. The second lens (the IR camera objective) is simplistically shown in Fig. 2, whereas its real structure may be much more complex.

Table 6. Operation modes of the THz objective

| Operation mode | Field of view | Magnification $M_1$ |
|---|---|---|
| A preliminary examination of the patient's skin area | 48 mm × 38.5 mm | 0.2 |
| Inspecting a chosen object | 9.6 mm × 7.7 mm | 1 |
| Inspecting details of the object | 1.9 mm × 1.5 mm | 5 |

It makes sense to assume the THz objective to be changeable, with different magnifications, say, $M_1 = 1; 0.2; 5$, whereby the two latter values can be obtained by inverting the same objective. The magnification $M_1 = 0.2$ would serve a preliminary examination of the patient's skin area, over the field of view (FOV) of about (9.6 mm × 7.7 mm)/0.2 = 48 mm × 38.5 mm (see Table 6). Operating with the magnification of $M_1 = 1$, one could inspect a chosen object within the FOV of 9.6 mm × 7.7 mm (for example, a pigmentary skin nevi). The magnification $M_1 = 5$ with its corresponding FOV of (9.6 mm × 7.7 mm)/5 ≈ 1.9 mm × 1.5 mm would enable inspecting details of the pigmentary skin nevi.

Thus, in our approach, the estimated length × width sizes of matrices of THz-to-IR converters are 9.6 mm × 7.7 mm.

The diffraction limits on the estimated objective's resolution are given in Table 7. At frequency 4.2 THz and $M_2 = 1$ (in the close-up operation mode) and pixel size $d = 15$ μm this means that the linear "uncertainty" of the image is covered, on the average, by (~71.3 μm)/(~15 μm) ≈ 5 pixels of the IR camera detector, that seems acceptable to resolve a meaningfully detailed pattern. At frequencies 0.38 and 8.4 THz (these are suitable, respectively, for contrast or sharp imaging of oncopathologies) and $M_2 = 1$, the linear "uncertainties" would be covered, on the average, by ≈26 and 2 pixels, respectively. If we correlate the image linear "uncertainty", expressed by the pixels number, with 512 pixels along the height of the IR camera detector (see Table 3), we get "uncertainty" as a percentage: 5 pixels – 1%, 26 pixels – 5%, 2 pixels – 0.4%; this appears to be an acceptable inaccuracy.

The lenses with desirable parameters (focal distance, diameter) for the THz range, made out of various materials, can be selected from the lists of commercially available products – see, e.g., Ref. 48.

The diffraction limit on the objective's resolution $\Delta x$ can be estimated as $\Delta x \approx \lambda \cdot l / A$ ($\lambda$: wavelength; $l$: distance between the objective centre and the object, hence patient's skin; $A$: the objective's aperture). The estimates show (see Table 7) that the enhanced frequencies, 4.2 and 8.4 THz, could yield acceptable resolutions for studying the structure of, say, pigmentary skin nevi.

Table 7. Operating parameters of the THz-to-IR converter with the 8.5 nm diameter GNPs. Estimations are done for $m_{el} = 4$, $n_{vm} = 1$ (see Subsec. 3.1.2 for details).

| Operating frequency (THz) | Wave-length (μm) | Photon energy (meV) | Photon momentum ($10^{-25}$ g·cm/s) | Objective's resolution (μm) at $l$ ~100 mm | Skin depth (nm) | $\delta p_{el}$ ($10^{-21}$ g·cm/s) | $\Delta p_F$ ($10^{-23}$ g·cm/s) |
|---|---|---|---|---|---|---|---|
| 0.38 | 789.5 | 1.57 | 0.84 | ~ 395.0[a] | 121.0 | | 1.8 |
| 4.2 | 71.3 | 17.4 | 9.3 | ~ 71.3[b] | 36.4 | ≥ 1.2 | 19.9 |
| 8.4 | 35.7 | 34.8 | 18.6 | ~ 35.7[b] | 25.7 | | 39.7 |

[a]$A$ ~200 mm; [b]$A$ ~100 mm

An operating frequency of 8.4 THz is acceptable if GNPs are used as NPs. If monel nanoclusters are used, the frequency can be even higher. From Fig. 2(a) in [49] (the DOSs of phonons over frequencies in Cu-Ni alloys with a broad peak in the vicinity of 5 THz), it follows that for the THz-to-IR converter with monel NPs, one can choose operating frequencies of 5 and 10 THz: there are many phonons with a frequency of 5 THz, while there are no phonons with a frequency of 10 THz in the monel NP. Therefore, the 10 THz photons could be absorbed due to the excitation of two phonons with a frequency of 5 THz. At the same time, the use of the 5 and 10 THz frequencies would provide an increased sharpness of the oncopathology image. Table 8 shows parameters of the THz-to-IR converter with the 2 nm diameter monel NPs.

Table 8. Operating parameters of the THz-to-IR converter with the 2 nm diameter monel NPs.

| Operating frequency (THz) | Wavelength (μm) | Photon energy (meV) | Photon momentum ($10^{-25}$ g·cm/s) | Objective's resolution (μm) at $l$ ~100 mm | Skin depth (nm) |
|---|---|---|---|---|---|
| 0.4 | 750 | 1.655 | 0.88 | ~ 375[a] | 560.8 |
| 5 | 60 | 20.7 | 11.0 | ~ 60[b] | 158.6 |
| 10 | 30 | 41.4 | 22.1 | ~ 30[b] | 112.1 |

[a]$A$ ~200 mm; [b]$A$ ~100 mm

We noted above that the same THz-to-IR converter could be used to capture the same area of skin at soft THz frequencies, that favors contrast enhancement, as well as at higher frequencies, that would enhance the image resolution. Comparison of images obtained at different frequencies would help to better define the margins of cancerous tissue.

Another approach is to use one THz-to-IR converter and two changeable IR cameras operating at different wavelengths, for example, Mirage and FLIR T650sc. This is possible if the converter's matrix material is transparent both at wavelengths of 1.5–5 μm and at ≈10 μm (for example, the Teflon® and polypropylene films).

## 3 Heating of metal NPs by THz radiation

Unfortunately, there is little data on the absorption of EM THz radiation by GNPs. Therefore, we turned to old works (1965–1978) on absorption by aluminum NPs in the THz frequency range [50-55]. For aluminum NPs this problem has been considered most consistently. As for the GNPs, there is a work [56] on measuring absorption in thiolate-protected gold clusters in the frequency range of ≈3–18 THz, which is in good agreement with the results of [50-55].

These experimental and theoretical studies demonstrate the promise of using metal NPs as nanotransducers of THz radiation into IR radiation (heat). We have identified the following physical mechanisms of THz radiation absorption, which may result in NPs' heating:

(a) direct transfer of the energy of EM THz radiation into mechanical energy of vibrations, e.g., into that of transverse phonons: transverse EM waves penetrating into a NP swing the lattice of ions if at resonance with the frequencies of transverse phonons;
(b) indirect transformation of the energy of EM THz radiation into mechanical energy of vibrations, e.g., into longitudinal phonons – via excitation of the Fermi electrons;
(c) simultaneous absorption of a THz photon and a primary longitudinal phonon by the Fermi electron, with a subsequent relaxation of an excited electron by generating a secondary longitudinal phonon whose energy equals the sum of the energies of a THz photon and a primary longitudinal phonon; the excited electron also could relax through multiple scattering by other electrons;
(d) dissipation of the energy acquired by the Fermi electron from a THz photon due to multiple scattering of an excited electron by other electrons; this physical phenomenon could be a part of the energy transformation processes (b) and (c) as a way for relaxation of the excited electron.

Although we ultimately turn to gold and monel NPs, rather than the aluminum ones, to be used in THz-to-IR converter, for the sake of better understanding the physical mechanism of NPs heating (a), we will first briefly review Refs. 50-55. The physical mechanism (b) was considered in Ref. 5, and a variety of realizations of the mechanism (c) in Ref. 57. Physical mechanism (d) takes effect if the energy level of an excited electron does not coincide with any of the energy levels of longitudinal phonons.

Experimental studies of the absorption of EM radiation by aluminum NPs in the frequency ranges of 0–1.5 THz [50] and 0–7.5 THz [51] have shown that the absorption coefficients of aluminum particles with sizes ~10–40 nm in the frequency range ~90 GHz–4.5 THz are proportional to the square of frequency.

These results were compared with the prediction of the Gor'kov-Eliashberg theory [52] for particles that took into account the quantization of the energy levels of electrons, as well as with the classical Drude theory. The Gor'kov-Eliashberg theory also predicts a quadratic dependence of the absorption coefficient on the frequency. In fact, the experimental results [50,51] demonstrated that the measured values of the absorption coefficient significantly exceed the predictions of the Gor'kov-Eliashberg theory.

Glick and Yorke have shown in their theoretical work [53] that strong absorption of small metal particles in the THz wavelength range was associated with the direct excitation of phonons due to the action of the EM wave field on the surface ions of the particle. It was noted that the authors of experimental works [50,51] were not able to explain the reason for such a large absorption, which by more than three orders of magnitude exceeded the predictions of both the Gor'kov-Eliashberg and the classical Drude theories. Glick and Yorke [53] for the first time pointed out an issue the authors of Refs. 50,51 did not pay attention to, and which might account for an unexpectedly strong absorption. They calculated the density of phonon states as a function of their frequency, compared their results with those of Ref. 54, and found a good match. Namely, the results of both studies had a quadratic dependence in the frequency range up to ~4 THz (this is the frequency range of transverse phonons). This correlation of the phonon state density profile and the absorption coefficient allowed Glick and Yorke to suggest that a strong absorption in aluminum NPs might be explained by direct excitation of transverse phonons by incident radiation. An explanation lacked, however, as to why the absorption coefficient increased with the particle size. A clue has been given by Granqvist [55] who anticipated, from theory considerations, an increase in the absorption coefficient for aluminum particles with their diameter, in the latter's range of 5–100 nm.

An immediate practical importance of experimental results [50,51,55] for the concept of THz-to-IR converter is that NPs of various sizes absorb in a broad range (see, e.g. Figs. 3,4 of Ref. 50), therefore an accurate selection of NP is not a critical issue (they all absorb anyway), and just setting some target spread of particle sizes would be acceptable. With this, the absorption is proportional to the size of the NPs.

Aluminum NPs cannot be used in the THz-to-IR converter because of their pyrophoricity. Therefore, we chose GNPs for the converter and used the data for aluminum by analogy. In relation to electronic properties at the Fermi level, with some simplification, gold can be considered as a simple metal. Both metals have the same crystal structure and almost identical lattice parameters. What's more important, GNPs are commercially available. We chose gold as a promising metal for NPs also because it does not oxidize at room temperature; therefore, GNPs surfaces are free of oxides that could absorb THz radiation.

Prerequisites for the absorption of EM radiation with the frequency $v$ by a metal NP of size $D$ due to the excitation of phonon with the frequency $v$ are the following:
– the NP's size $D$ is smaller than the penetration depth $d$ of EM radiation with the frequency $v$: $D \leq d$;
– the phonon wavelength $\lambda$ at the frequency of the incident EM radiation $v$ is not larger than the NP's size $D$: $\lambda \leq D \leq d$;
– the Heisenberg relation for the momenta of a phonon and an electron in a NP of size $D$ ensures the fulfillment of the law of conservation of momentum upon excitation of a phonon;
– the GNP's size $D$ should not be smaller than the size at which spontaneous emission of THz radiation by a (small enough) GNP begins, i.e., $D > 8$ nm [58], in order to prevent the GNP's cooling via emission of THz photons. This observation was guided by our study [58] dedicated to a suggested explanation of the size effect in the heterogeneous catalysis on GNPs (see also Appendix A in this paper). Nanoparticles with sizes inferior to 8 nm tend to cool down due to a spontaneous emission of THz photons, hence become useless if the objective is opposite, to heat the GNPs in the THz-to-IR converter by incoming THz radiation. On the other side, an increased *D* would degrade the converter's sensitivity, since an elevated THz power level will be required to efficiently heat the GNPs.

From the results of Refs. 50-55 for aluminum NPs it follows that, if the latter satisfy the aforementioned prerequisites, an absorption of EM radiation takes place at all wavelengths at which phonons, whether transverse or longitudinal, exist in a NP. According to the phonon dispersions along the Γ–X high symmetry direction in the space of wave vectors in gold [59], the maximum frequencies of longitudinal and transverse phonons are ≈4.5 and ≈2.55 THz, respectively. For such EM radiation frequencies, the skin depths in gold are 35.2 and 46.7 nm, accordingly. On the other hand, we would like to operate GNPs at frequencies 4.2 and 8.4 THz (see Table 7), for which the skin depths are 36.4 and 25.7 nm, respectively. As for the frequency of 0.38 THz, suitable for visualization of oncopathologies, the skin depth for it is ≈121 nm (see Table 7). Hence, the 8.5 nm diameter GNPs are transparent for all important EM radiation frequencies.

The size of NPs in the THz-to-IR converter is much smaller than the wavelength of the incident THz radiation. However, the Rayleigh scattering does not occur, since the aforementioned prerequisites for the absorption of EM radiation, $\lambda \leq D \leq d$, are satisfied. Under these conditions, THz radiation will not be scattered into the solid angle of $4\pi$, but will penetrate into the NP, in which the physical mechanisms of heating (a)–(d) identified above will set to work.

Now let us consider in more detail the energy and momentum conservation relations in physical mechanisms (a)–(d) of THz radiation absorption.

## 3.1 GNPs

### 3.1.1 *Direct transfer of the energy of THz radiation into mechanical energy of vibrations (transverse phonons)*

Absorption of EM radiation in the frequency range of transverse phonons in aluminum can be explained based on the assumption of direct interaction of incident photons with transverse phonons [53,54]. Arguably, a similar direct interaction takes place in GNPs. Figure 6 demonstrates how the conservation laws are fulfilled in the interaction of an EM photon and a transverse phonon. Dispersion curves of photons and transverse phonons do not intersect, since the speed of phonon propagation in gold is much less than the speed of light. How does a photon directly excite a transverse phonon in a NP, as suggested by Glick and Yorke [53]? The explanation lies in confinement: due to the uncertainty in the momentum of transverse phonon $\Delta p$, there is a possibility of its excitation by a photon. The Heisenberg uncertainty relation for momentum and coordinate reads:

$$\Delta p \cdot \Delta x \geq \hbar. \qquad (1)$$

Let us express $\Delta p$ through the uncertainty in the wave number $\Delta k$ : $\Delta p = \hbar \cdot \Delta k$ and assume $\Delta x = D$; then the Heisenberg relation (1) takes the form: $\Delta k \geq (1/D)$. Thus, transverse phonons in a NP of diameter $D$ have an uncertainty in the wave number $\Delta k$ not less than $(1/D)$. For 8.5 nm diameter GNPs, $\Delta k_{min} \approx 1.18 \cdot 10^6$ cm$^{-1}$.

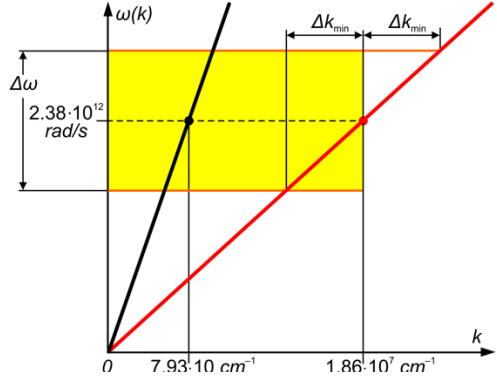

Fig. 6 (arbitrary scale). Explaining the excitation of a transverse phonon by a photon in NP. Black and red straight lines are the dispersion laws of photons and transverse phonons, respectively. Due to the uncertainty in the wave number $\Delta k$ of transverse phonons, the gap between the wave numbers of the photon and the phonon overlaps, and there is a transfer of energy and momentum from the photon to the transverse phonon.
All numerical data are for the frequency of 0.38 THz ($\omega = 2.38 \cdot 10^{12}$ rad/s).

Let us choose a frequency from the range of 0.35–0.55 THz, acceptable for visualization of the human skin oncopathologies, for example, 0.38 THz ($\omega = 2.38 \cdot 10^{12}$ rad·s$^{-1}$). This frequency is also chosen because it satisfies conditions (5) and (6) in Subsec. 3.1.2 below. These photons have the wave number $k = 7.93 \cdot 10$ cm$^{-1}$. The attenuation of the 0.38 THz radiation passing a 1 m distance in air (this is the characteristic distance between the THz radiation source and the IR camera in the proposed method, see Fig. 2) is 0.06 dB [60]. That means that the THz radiation power is attenuated by a factor of approximately 1.014, which can be neglected.

From the velocity $v_T = 1.28 \cdot 10^5$ cm·s$^{-1}$ of transverse phonons along the Γ–X direction, we determine their wave number $k$, corresponding to the frequency of 0.38 THz: $k = \omega/v_T = 1.86 \cdot 10^7$ cm$^{-1}$.

Thus, in gold, at the frequency of 0.38 THz, the gap in the wave numbers of photons and phonons is enormous (see Fig. 6): these values differ by almost 5 orders of magnitude. The gap exceeds the value of $\Delta k_{min}$ by a factor of about 16. At first glance, at the frequency of 0.38 THz, the interaction between a photon and a transverse phonon is impossible (as if a photon would not be able to excite a transverse phonon). But in NPs, due to confinement, the uncertainty in the wave number of transverse phonons obviously allows the interaction of photons and transverse phonons. The transverse electric field of the electromagnetic wave swings the metal ions in the NP, and this takes on a semblance of a transverse phonon. Otherwise, it is difficult to explain the experimentally observed [50,51] far-IR absorption in small metallic particles, namely, at the frequencies of excitation of transverse phonons.

Therefore, we argue that the excitation of transverse phonons by THz photons contributes to heating the NP. Under this angle, the 8.5 nm diameter GNPs are an acceptable choice for the THz-to-IR converter.

According to the phonon dispersions in the space of wave vectors in gold [59], the propagation velocity of transverse phonons for the Γ–K direction is maximal and greater than that for the Γ–X direction (respectively, $1.45 \cdot 10^5$ cm·s$^{-1}$ and $1.28 \cdot 10^5$ cm·s$^{-1}$). However, for the Γ–K direction, the gap in the wave numbers of photons and phonons is large too: it exceeds the value of $\Delta k_{min}$ by a factor of 14.

Let us now estimate the minimum frequency of transverse phonons capable of being excited in a GNP with a diameter $D$. Under an assumption that the phonon dispersion relations in GNP do not differ much from those in bulk gold [59], the propagation velocity of transverse phonons can be estimated from the latter. The wavelength of transverse phonons is $\lambda_T = (v_T/\nu)$, where $v_T$ is the propagation velocity of transverse phonons. Since the maximum wavelength that can be excited in a NP is equal to the diameter $D$, the minimum frequency of transverse phonons that can be excited in a NP is $\nu_T^{(min)} = (v_T/D)$. This directly transposes into the minimum frequency of EM THz radiation absorbable due to the direct excitation of transverse phonons. So, for the 8.5 nm diameter GNP, assuming for simplicity the Γ–X propagation direction, we arrive at $\nu_T^{(min)} =$ 0.15 THz, using the respective Γ–X propagation velocity of transverse phonons of $v_T = 1.28\cdot 10^5$ cm·s$^{-1}$. Thus, the 8.5 nm diameter is acceptable in terms of being able to heat GNPs by transverse phonons.

### 3.1.2 *Indirect transformation of the energy of THz radiation into mechanical energy of vibrations (longitudinal phonons)*

We come now to discussion of the physical mechanism for heating the GNPs by indirect converting the THz to IR via excitation of the Fermi electrons. First, let us consider a simplified situation in GNP, when the energy interval between the Fermi level and a nearby accessible vacant electron energy level equals the energy of the THz photon, $h\nu$. Figure 7 depicts a case of absorption of a THz photon by a Fermi electron; the subsequent relaxation of the excited electron releases a longitudinal phonon. The surface of revolution around the vertical (energy) axis in Fig. 7(b) schematically shows the energy dispersion of free electrons as function of just two momentum coordinates (here the notation $p_{x,y,z}$ should be understood so that in the plane $E = E_F$ there can be pairs of electron momentum components: $p_x$ and $p_y$ or $p_x$ and $p_z$ or $p_y$ and $p_z$). The electron energy dispersion surface is delimited at the bottom by Fermi momentum/energy and at the top – by a discrete energy level accessible following an excitation. The smearing in the upper momentum plane indicates the uncertainty of the electron momentum due to spacial confinement (see below), that helps to match the momentum conservation condition involving a photon.

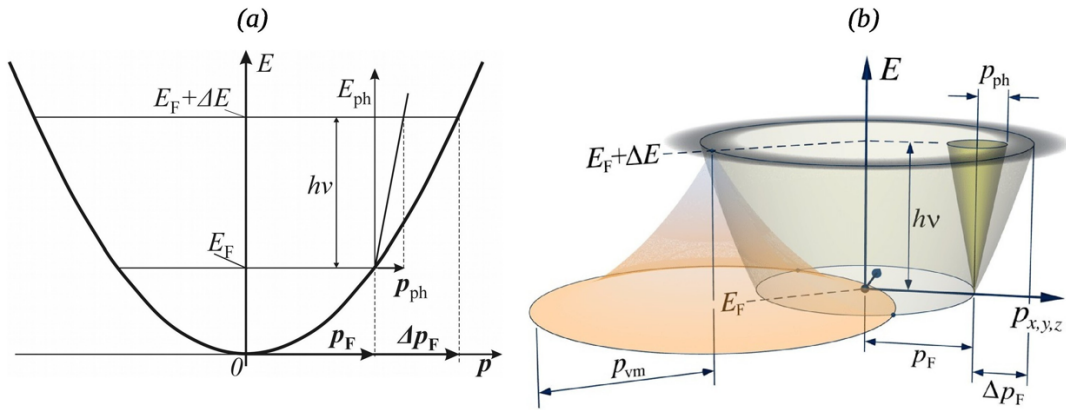

Fig. 7. Absorption of THz photon by GNP helped by uncertainty of the Fermi electron's momentum.
*(a)* Interaction of a THz photon of energy $h\nu$ with a Fermi electron. Here $\Delta E = m_{el}\cdot\Delta E_{el} = h\nu$.
*(b)* Right part of the figure illustrates absorption, while its left part depicts subsequent relaxation via releasing a phonon. The dispersion law of an electron is depicted by ascending paraboloid, that of the photon – by narrow ascending cone, that of longitudinal phonon – by descending bell mouth. Dispersions are shown over two-dimensional momentum plane. The crossing points of the electron and phonon dispersion surfaces at $E = E_F$, marked by dots, indicate relaxation points for the excited electron. See text for details.

The dispersion of a THz photon is shown by a narrow cone, starting from some momentum/energy value of a Fermi electron. In the follow-up of an excitation to the energy $E_F + \Delta E$, the electron can relax releasing a longitudinal phonon. The phonon dispersion is depicted by a descending (inverted) bell mouth, starting from some momentum/energy of the excited electron. Essential observations are that (i) for small enough NPs, the momentum and energy conservation in the course of an immediate electron excitation by a THz photon can be assured by the uncertainty relation; (ii) following the relaxation of an excited electron via releasing a phonon, the electron's momentum may undergo a large reorientation (largely retaining its magnitude).

Let us now consider some details more attentively, in the general case with an arbitrary number of electron energy steps $m_{el}$ and number of energy steps $n_{vm}$ of the vibration modes.

First, we try to estimate the sizes of GNPs which could be efficiently used in the THz-to-IR converter. The NP diameter $D$ allows to "tune" the electron energy separation $\Delta E_{el}$, according to the Kubo formula for the level spacing in a NP, [61,62] (see also Appendix 1 in Ref. 57). The energy delivered by a THz photon $h\nu$ should match an integer number of steps, $m_{el}$, in the electron excitation ladder; simultaneously it should fit an integer number $n_{vm}$ of energy steps $\Delta E_{vm}$ for the vibration mode:

$$m_{el}\, \Delta E_{el} = h\nu = n_{vm}\, \Delta E_{vm} \,. \qquad (2)$$

For simplicity and order-of-magnitude estimates, we assume a linear dispersion law for the longitudinal phonons, supposing that they are propagating along the linear size for which we take the diameter $D$, possessing the nominal sound velocity $v_L$, hence $\Delta E_{vm} = v_L h/D$. A reference to the Kubo formula modifies Eq. (2) as follows:

$$m_{el}\cdot(4/3)\cdot(E_F/N) = h\nu = n_{vm}\cdot v_L\cdot(h/D) \,, \qquad (3)$$

where $E_F$ is the Fermi energy of gold, $N$ is the number of gold atoms in the nanosphere, that is otherwise the ratio of the nanosphere's volume $V = (4/3)\cdot\pi\cdot(D/2)^3$ to the volume per atom in the gold's face-centered cubic lattice (with the lattice constant $a_{Au}$), $V_{at} = a^3_{Au}/4$, hence $N = V/V_{at} \approx 2.09\cdot(D/a_{Au})^3$. With this, the diameter $D$ is expressed via the $m_{el}$ and $n_{vm}$ parameters as follows:

$$D \approx 0.798\cdot a^{3/2}_{Au}\cdot[(m_{el}/n_{vm})\cdot(E_F/v_L h)]^{1/2}, \qquad (4)$$

or, with the material constants inserted and specifying the units,

$$D \approx 4.23\cdot(m_{el}/n_{vm})^{1/2}\ nm. \qquad (5)$$

It follows from the right parts of Eqs. (2) and (3) that

$$\nu = (n_{vm}\, \Delta E_{vm})/h = n_{vm}(v_L/D) \,. \qquad (6)$$

Accepting the nominal value of the velocity of sound in gold $v_L = 3.23\cdot10^5$ cm·s$^{-1}$, we arrive at the following parameter values: $\nu = 0.38$ THz, $D \approx 8.5$ nm; $m_{el} = 4$; $n_{vm} = 1$.

In addition to the above arguments based on the energy conservation law, we should pay attention to issues of the momentum conservation. On excitation of a Fermi electron by a photon of energy $h\nu$, the electron momentum (assuming a free-electron dispersion law) gets an increase $\Delta p_F \approx h\nu/v_F$, where $v_F \approx 1.4 \cdot 10^8$ cm·s$^{-1}$ is the Fermi velocity of electrons in gold [63]; hence, at $\nu$ = 0.38 THz, $\Delta p_F \approx 1.8 \cdot 10^{-23}$ g·cm·s$^{-1}$.

The momentum of the absorbed THz photon, $p_{ph}$, is much smaller than $\Delta p_F$ and therefore does not ensure the fulfillment of the momenta conservation law. However, at small enough GNP sizes, the value of $\Delta p_F$ might be absorbed by the uncertainty in the electron's momentum $\delta p_{el}$. Indeed, according to the Heisenberg's uncertainty relation, in a confined geometry of the GNP, the uncertainty of the electron's momentum reads $\delta p_{el} \geq h/(2\pi D)$. For the GNP size we consider, $D$ = 8.5 nm, $\delta p_{el} \geq 1.2 \cdot 10^{-21}$ g·cm·s$^{-1}$. Hence, the condition $\delta p_{el} \geq \Delta p_F$ is by far respected. Therefore, the mismatch of the electron momenum to satisfy the momenta conservation on absorbing a THz quantum with frequency $\nu$ = 0.38 THz would be "absorbed" by the Heisenberg uncertainty relation.

According to Eqs. (5) and (6), the values of $D$ and $\nu$ will be discrete corresponding to the integer values of $m_{el}$ and $n_{vm}$. It could be assumed that the NP would exhibit "resonant" absorption at frequencies $\nu$ when tuned to frequency $\nu$ = 0.38 THz. Similarly, one could assume that for a given frequency $\nu$ = 0.38 THz, the chosen diameter $D$ = 8.5 nm would also be "resonant".

However, real NPs do not exhibit resonant absorption properties either in $D$ or in $\nu$. In any case, the experimental results for aluminum nanoparticles (Figures 3, 4 of Ref. 50) show no absorption selectivity either in diameter or in frequency. It could be assumed that this is a consequence of the variation in the shape and size of NPs – due to which the quantization for different directions and sizes of NPs varies, and in absorption experiments – the absorption coefficient is averaged over many NPs.

We assume that the absence of resonances neither in $D$ nor in $\nu$ is explained by other properties of the NPs. We are faced with a paradox: on the one hand, the small size of the particle implies discretization of energy levels and quantization of the momenta of electrons and phonons, consequently, the "resonant" diameter of the particle. On the other hand, just because of the smallness of the particle, due to the confinement of electrons and phonons in it, there is uncertainty of their energy and momenta levels. That is, the sharpness of the criteria for the energy levels within a NP to be discrete is somehow dissolved. Therefore, photon absorption may occur not necessarily at integer values of the number of steps in energy and momenta.

Let us now calculate the minimum frequency of longitudinal phonons that can be excited in a GNP with a diameter $D$. We again suppose that the phonon dispersion curve in GNP is close to that in bulk gold, and we estimate the propagation velocity of longitudinal phonons from the phonon dispersion curve in bulk gold [59]. The wavelength of longitudinal phonon is $\lambda_L = (v_L/\nu)$, where $v_L$ is the propagation velocity of longitudinal phonons. The maximum wavelength excited in a NP is equal to the NP's diameter $D$; therefore, the minimum frequency of the longitudinal phonon excited in the NP is $\nu_L^{(min)} = (v_L/D)$. Hence, the minimum frequency of EM THz radiation, which can be absorbed due to the indirect excitation of a longitudinal phonon through the excitation of an electron, is also $\nu_L^{(min)}$.

Assuming for simplicity that the direction of propagation of longitudinal phonons in a GNP coincides with the high symmetry direction Γ–X, and using the propagation velocity of longitudinal phonons along the direction Γ–X, $v_L = 3.23 \cdot 10^5$ cm·s$^{-1}$, for a 8.5 nm diameter GNP, we get the minimum frequency $\nu_L^{(min)}$ = 0.38 THz. Thus, the GNP's diameter of 8.5 nm is acceptable for heating the GNP due to the excitation of longitudinal phonons by EM radiation with a "medical" frequency of 0.38 THz.

According to the theoretical calculations (Refs. 64-66), in GNPs, the vibrational (phonon) DOS in the frequency range 0.35–0.55 THz is low. However, experimental studies reveal the Lamb modes in GNPs precisely in the range of "medical" frequencies (see Fig. 5 in Ref. 64), which would be valuable for the implementation of our idea of visualization of oncopathologies.

Combining the findings of Subsecs. 3.1.1 and 3.1.2, we conclude that a GNP with a diameter of 8.5 nm is suitable for absorbing the "medical" frequency of 0.38 THz. Moreover, for higher frequencies of THz radiation, it is capable of being heated by transverse and longitudinal phonons, and can be used as a nanoconverter of THz radiation into IR radiation.

### 3.1.3 *Simultaneous absorption of a THz photon and a longitudinal phonon by the Fermi electron*

An increase in the efficiency of absorption of 0.38 THz photons may occur due to simultaneous absorption of a 0.38 THz photon and a longitudinal phonon by a Fermi electron. This can be realized due to the uncertainty in the Fermi energy when a NP tuned by its diameter to absorb the 0.38 THz photon indirectly (namely, a 8.5 nm diameter GNP with the chosen parameters $m_{el} = 4$ and $n_{vm} = 1$, see Subsec. 3.1.2) turns out capable to absorb a THz photon of the same frequency with the participation of a longitudinal phonon. To make this particular absorption important, the energy of the longitudinal phonon concerned should fall within the full width at half maximum (FWHM$_L$) of the peak of the energy distribution of longitudinal phonons. The relaxation of the excited electron will occur due to the excitation of the secondary longitudinal phonon.

Figure 8 shows two particular cases of simultaneous absorption of the THz photon and longitudinal phonon by each of the two excited Fermi electrons, whereby the electrons are excited from possible states ($E_F - h\nu$) and ($E_F + h\nu$) (these states are within the uncertainty zone of the Fermi level and within the range of thermal smearing $k_B T$, see Appendix B, Fig. 14). Figure 8 assumes the incidence of photons from a fixed direction, therefore, the photon momenta $p_{ph}$ in Fig. 8 are fixed for this direction. Two momenta of the Fermi electrons are considered, coinciding with the direction of the photon momentum and opposite to it. In Fig. 8, black cones are surfaces of rotation of a straight line, the law of dispersion of photons, and the bell-shaped red surfaces are surfaces of rotation of a generatrix, the law of dispersion of longitudinal phonons in gold.

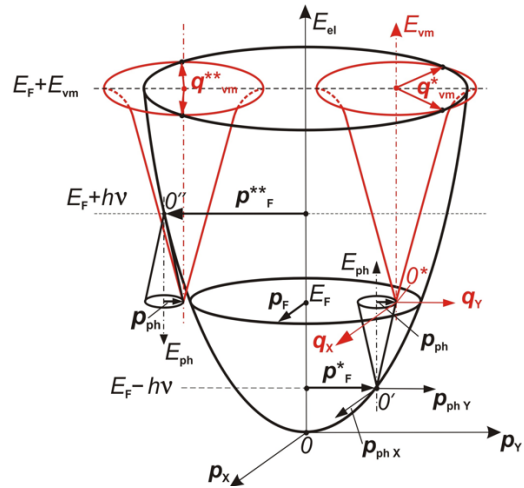

Fig. 8. Scheme of a simultaneous absorption of a THz photon with energy $h\nu$ and a longitudinal phonon with energy $E_{vm}$ by the Fermi electron in GNP (arbitrary scale).

The intersection points of the red surfaces and the paraboloid of revolution (the dispersion surface of electrons) in the plane $E = (E_F + E_{vm})$ fix the pairs of momenta $\boldsymbol{q}^*_{vm}$ and $\boldsymbol{q}^{**}_{vm}$, which are transferred by longitudinal phonons with energy $E_{vm}$ to electrons.

Figure 8 conventionally depicts the situation when the energy of the Fermi electron is not defined: it can be attributed to one of the incompletely occupied levels in the range from $[E'_F - (k_B T/2)]$ to $[E''_F + (k_B T/2)]$ (see Fig. 14, Appendix B). On the right is the case when the electron momentum $\boldsymbol{p}^*_F$ in the possible state $(E_F - h\nu)$ and the photon momentum $\boldsymbol{p}_{ph}$ are directed in the same direction. On the left is the case when the electron momentum $\boldsymbol{p}^{**}_F$ in the possible state $(E_F + h\nu)$ is opposite to the photon momentum $\boldsymbol{p}_{ph}$.

For simplicity, in Fig. 8, a special case is shown when the point $0^*$, the origin of coordinates for the dispersion surface of longitudinal phonons, coincides with the end of the vector $\boldsymbol{p}_{ph}$. But due to confinement, point $0^*$ is located in the smear area, within the ring of uncertainty of the electron momentum at the $E_F$ level, similar to the ring of uncertainty shown in Fig. 7(b).

From the data on the vibrational density of states in gold nanocrystals, Refs. 64-66, one can estimate that in GNPs, the FWHM$_L$ band lies within ≈3.7–4.6 THz. This means that in the simultaneous absorption of 0.38 THz photons according to the scheme in Fig. 8, the longitudinal phonons from the low-frequency half of the FWHM$_L$ band (with frequencies ≈3.7–4.22 THz) could participate most effectively. They would be able to excite electrons to levels corresponding to frequencies of ≈ 4.08–4.6 THz. The excited electrons would be able to relax, exciting longitudinal phonons from the high frequency half of the FWHM$_L$ band (with frequencies ≈4.28–4.6 THz). Each of the excited longitudinal phonons could redistribute energy between the secondary longitudinal phonon of lower energy and the electron with energy within a zone of width $\delta E_{el}$ (see Fig. 15, Appendix B), similar to mechanism considered in Ref. 67. Dissipation would proceed along the mixed dissipation, electron–phonon scattering, with the participation of longitudinal phonons.

The participation of dominant phonons (from within the FWHM$_L$ band) in this process increases a probability of absorption of 0.38 THz photons according to Fig. 8, where a special case of simultaneous absorption is shown with electron states $(E_F - h\nu)$ and $(E_F + h\nu)$, with $\nu = 0.38$ THz.

Summarizing, we formulate the properties of NPs, the combination of which would both provide the most efficient conversion of the energy of THz radiation into heat and ensure convenient operational capabilities of the THz-to-IR converter:
(1) transparency of NPs in a wide frequency range (larger skin depth than the size of NPs);
(2) the perception of photon energy by electrons in a wide EM radiation frequency range (due to confinement and uncertainty in the energy of electron energy levels);
(3) fulfillment of the condition of energy dissipating in the NP, that is, the condition of relaxation of the excited electron inside the NP's volume due to electron-electron and electron-phonon scattering (then the energy of the excited electron remains in the NP and is not carried away from it in the form of photon emission): $l_{mfp} < (D/2)$; hence, the requirement for the particle diameter is $D > 2 l_{mfp}$.

The third property in 8.5 nm diameter GNPs is obviously not fulfilled: its implementation would mean that $l_{mfp} < 4.25$ nm in these GNPs. But an analysis of the behavior of $l_{mfp}$ in gold alloys versus their composition shows (see Table 9) that this is hardly possible in pure gold. Indirect confirmation that $l_{mfp} > 8$ nm in GNPs follows from the fact that catalysis on GNPs stops at GNPs diameters greater than ~8 nm (see Appendix A).

As already mentioned, the phonon DOS of gold in the frequency range 0.35–0.55 THz is low. Therefore, it makes sense to consider one more way to increase the efficiency of NP heating by THz radiation of these frequencies, namely, to switch to a NP with $l_{mfp}$ smaller than the NP's radius. Then the NP will heat up predominantly due to the physical mechanism (d) "dissipation of the energy acquired by the Fermi electron from a THz photon due to multiple scattering of an excited electron by other electrons" (see the beginning of the Sec. 3). That is, since there are few phonons, we should address the advantage of intense electron-electron scattering.

We considered ways to reduce the value of $l_{mfp}$ by alloying gold with palladium. As we stated earlier, gold was selected as a material for the NPs because it does not oxidize at room temperature; hence, there are no oxides capable of absorbing THz radiation on the GNP surface. Palladium, like gold, does not oxidize in air at room temperature. Palladium possesses peak of electron DOS at the Fermi level [68] and forms continuous row of solid solutions with gold [69]. The alloys with compositions close to ~50at.%Au50at.%Pd seemed promising in the sense of being the most distorted, because they were "as much as possible away" from lattices of pure constituents. The distortion induces intense scattering of electrons. The de Broglie wavelength of the Fermi electrons is ~0.5 nm. This seemed to be reasonably commensurate with the expected areas with distorted periodicity in the alloy made out of Au and Pd, characterized by lattice parameters of 0.408 and 0.389 nm, respectively [69].

The following additional observations summarized the arguments as for the "usefulness" of alloying:

(i) As compared to pure gold, the Au-Pd alloys possess enhanced electron DOS at the Fermi level [70-72]; in these conditions, due to a thermal smearing of the DOS distribution, the electron states near the Fermi level are occupied partially. Therefore, in Au-Pd alloys, an enhanced number of unoccupied electron states close by the Fermi level causes an intensive scattering of excited electrons.

(ii) The resistivity of the Au-Pd alloys at compositions ~50at.%Au50at.%Pd reaches maximum values [73,74], an order of magnitude higher than for pure gold.

However, it turned out that even such promising prerequisites do not allow to significantly reduce the value of $l_{mfp}$ even in the high resistivity alloy 50at.%Au50at.%Pd: according to our estimates, its $l_{mfp}$ is ≈7.8 nm (see Table 9). The diameter of NPs corresponding to this $l_{mfp}$ would be quite large: $D \approx 16$ nm. This would not allow to obtain acceptable parameters for the converter (see evaluations in Table 1).

Introducing Ta or Fe impurity atoms into gold provides a peak in the electron DOS at the Fermi level [75]. However, this approach, i.e., using the Ta or Fe impurity atoms, also showed that $l_{mfp}$ would be large and the diameter of such particles would be large too; accordingly, this will result in unacceptably high values of the volume to be heated and for the power required to heat the NP to be seen by the IR camera.

Fe impurities in Au form $d$-levels of electrons at the Fermi level of gold [75]. With iron impurities in Au up to 3.2at.%Fe, the resistivity increases to 28 µOhm·cm [73]. The Au-23at.%Fe alloy has ≈3.4 times greater resistivity, 95 µOhm·cm [76]. It is possible that this alloy too has a peak in the electron DOS at the Fermi level.

Using the Drude model, we roughly estimated the expected values of $l_{mfp}$ in the Au-3.2at.%Fe and 77at.%Au23at.%Fe alloys from the available resistivity data (see Table 9). It is unlikely that one would manage to significantly reduce $l_{mfp}$ in gold (or gold alloy) NPs.

Thus, in bulk gold, the mean free path of electrons at 300 K is fairly large: $l_{bulk} \approx 37.7$ nm [77] and it is not easy to significantly reduce it.

Table 9. The electron mean free path $l_{\text{mfp}}$ in gold-based alloys

| Gold-based alloy | $\rho$ (µOhm·cm) | $l_{\text{mfp}}$ (nm) | Ref. |
|---|---|---|---|
| 50at.%Au50at.%Pd | 20.57 | 7.8 | [78] |
| Au-2.2at.%Ti | 30 | ≈ 5.6 | [73] |
| Au-3.2at.%Fe | 28 | ~6 | [75] |
| 77at.%Au23at.%Fe | 95 | ~2 | [76] |

Unfortunately, we did not find data (density, specific heat and thermal conductivity) for Au-2.2at.%Ti, Au-3.2at.%Fe and 77at.%Au23at.%Fe alloys to estimate their parameters using the heat equation. We could estimate the required data only for the 50at.%Au50at.%Pd alloy and calculated its parameters (see Table 1).

Thus, recognizing that all physical mechanisms (a)–(d) and prerequisites for the absorption of EM radiation, listed in the beginning of the Sec. 3, are implemented in the 8.5 nm diameter GNP, we note that the mechanism related with the requirement $D \geq 2l_{\text{mfp}}$ is not implemented in GNP. Under confinement conditions, it could become synergistic if combined with the listed mechanisms (a)–(d), but, unfortunately, this did not happen.

From the comparative graph in Fig. 5 in Ref. 79, it can be seen that from Au-Ni, Cu-Ni and Ag-Pd alloys, which form solid solutions with a peak in the electron DOS at the Fermi level and therefore have a resistivity peak on the resistivity-composition curve, it is more advantageous to choose a Cu-Ni alloy, because it has a higher resistivity, and, therefore, can provide a shorter $l_{\text{mfp}}$. Obviously, gold-based alloys will have to be abandoned in favor of the Cu-Ni alloy.

### 3.2 *Copper-nickel MONEL® alloy 404 NPs*

In this Subsection, we will consider an analogue of the 50at.%Au50at.%Pd alloy – monel 47wt%Cu53wt%Ni, in which, due to the states of copper *d*-electrons at the Fermi level of nickel, it is possible to increase the electron DOS in a band ~ ($k_\text{B}T$/2) wide below $E_\text{F}$ to such an extent that a high resistivity be achieved (~49.8 µOhm·cm), and $l_{\text{mfp}}$ to become as small as ≈0.85 nm.

The 47wt.%Cu53wt.%Ni alloy is very close in composition to the commercial MONEL® alloy 404 UNS N04404 (see Ref. 80), whose data (density, specific heat and thermal conductivity) are known, allowing the required parameters to be calculated using Eq. (7) in Sec. 5. Therefore, in Tables 10 and 11 and the graphs in Fig. 11, for definiteness, the grade of this commercial alloy is indicated, although the term "monel" is used in the text for simplicity.

Differently from the case of the 50at.%Au50at.%Pd alloy, in the monel the synergy of three important properties can be realized: transparency of NPs in a wide frequency range of EM radiation, the perception of photon energy by electrons in a wide EM radiation frequency range, and fulfillment of the condition of energy dissipating in the NP. Let us elaborate.

For frequencies 0.35–0.55 THz, the skin depth in monel alloy is $d ≈ 602–480$ nm. This exceeds the 2 nm diameter of monel NPs by more than two orders of magnitude. Therefore, monel NPs are transparent for frequencies favorable for imaging in oncology ("medical" frequencies), $\nu$ = 0.35–0.55 THz.

It should be noted that for the THz-to-IR converter with monel NPs, instead of the operating frequency of 0.38 THz we can choose 0.4 THz – at this frequency, the attenuation in air is slightly lower and equals 0.05 dB/m [60], or 1.012 times at a distance of 1 m, which can be neglected. We recall that the frequency of 0.38 THz was chosen for GNPs based on a consideration that the numbers $m_{\text{el}}$ and $n_{\text{vm}}$ must be integers. However, for NPs of the monel alloy this is not required, and one can choose a more acceptable 0.4 THz instead of the operating frequency of 0.38 THz.

In general, estimates show that a NP of the monel alloy 2 nm in diameter is transparent at 0.1–10 THz with photon energies 41–0.41 meV. Therefore, photons of the entire THz range can penetrate the NP and interact with Fermi electrons, exciting them to states with energies 0.41–41 meV above the Fermi level. Consequently, due to electron-electron scattering, such a NP is capable of converting not only the "medical" THz radiation into heat, but also the operating radiation with frequencies of 5 and 10 THz (see Table 8). Note that the latter can also be absorbed due to the excitation of longitudinal phonons, since they exceed the minimum frequency $v_L^{min}$ = 2.15 THz in a 2 nm monel NP, see below, which could provide a sharper imaging of the human skin oncopathology.

In a search for verification of the energy dissipation condition in a monel NP, we turned to an article that combines the $l_{mfp}$ estimates for copper-nickel alloys done by four groups of researchers [81]. According to this article, the monel alloy has a very favorable, for our purposes, $l_{mfp}$ value, as small as ≈ 0.85 nm. In accordance with the requirement $D \geq 2l_{mfp}$, the diameter of the monel NP was chosen: $D$ = 2 nm. Then an excited electron, having passed further a path equal to $l_{mfp}$, will relax by electron-electron scattering. So, a multiple scattering of electrons with energy dissipation takes place inside the 2 nm monel NP, which is what we wanted.

It is useful to compare the capabilities of NPs from nickel and the copper-nickel alloys of monel and constantan. The analysis of data on the electron DOS at the Fermi level $N(E_F)$ of $Cu_x$-$Ni_{1-x}$ alloys (see Fig. 2 in Ref. 81) together with the dependence of the resistivity of copper-nickel alloys on the content of components (see, for example, Fig. 12.5 in Ref. 82) and the results of $l_{mfp}$ estimates for copper-nickel alloys [81] shows that Ni has a non-optimal combination of an increased $N(E_F)$ and a large $l_{mfp}$. Therefore, although the number of electrons in Ni which would be able to absorb the energy of THz radiation is large, the intensity of their scattering is not high enough. Consequently, the efficiency of conversion of THz radiation into heat is not high either. On the contrary, for the monel alloy, although the $N(E_F)$ is ≈3 times lower than that of pure Ni, $l_{mfp}$ is minimal among copper–nickel alloys and by ≈8 times shorter than that of Ni [81]. As a result, the scattering intensity of an excited electron in the monel alloy is ≈8 times higher than that of Ni. That means, the conversion efficiency of the monel alloy is expected to be ~2.7 times higher than that of pure Ni.

According to the graph in Fig. 2 in Ref. 82, monel has an advantage over constantan too: its $N(E_F)$ is larger, therefore, the efficiency of conversion of THz radiation into heat is higher. Moreover, according to the graph in Fig. 12.5 in Ref. 83, monel has a higher resistivity than constantan, hence higher scattering intensity and shorter $l_{mfp}$. Combining these two considerations, we conclude that monel is preferable to constantan.

Thus, of the three compared metals for NPs, monel can provide the best dissipation of THz radiation energy. As noted above, in Sec. 2, due to the low $l_{mfp}$ in the monel alloy, the NP diameter $D$ can be chosen small, which reduces the NP's volume to be heated by THz photons and, consequently, increases the sensitivity of the THz-to-IR converter in terms of THz power.

We estimated the propagation velocities of transverse $v_T$ and longitudinal phonons $v_L$ in monel from the dispersion curves Fig. 4b) in Ref. 84. They turned out to be, respectively, $v_T$=3.3·10$^5$ cm·s$^{-1}$ and $v_L$=4.3·10$^5$ cm·s$^{-1}$. This made it possible to estimate the minimum phonon frequencies that can be excited in a monel NP with a diameter of 2 nm (assuming that the maximum wavelength that can be excited in a NP equals the latter's diameter $D$): for transverse phonons $v_T^{min} = (v_T/D)$ = 1.65 THz, and for longitudinal phonons $v_L^{min} = (v_L/D)$ = 2.15 THz.

Thus, the minimum frequency of EM radiation that can be directly absorbed by a monel NP with a diameter of 2 nm due to the excitation of a transverse phonon is 1.65 THz. And indirectly, through the mediation of an excited Fermi electron, the same NP will be able to absorb EM radiation with a minimum frequency of 2.15 THz.

Hence it follows that the "medical" frequencies 0.35–0.55 THz cannot be absorbed due to the excitation of transverse phonons: they cannot swing transverse phonons under the action of the electric vector of the EM wave, since the "medical" frequencies are less than $v_T^{min}$. A photon of EM radiation of "medical" frequencies can excite an electron in a 2 nm diameter monel NP, but an excited electron will not be able to further excite a longitudinal phonon, since "medical" frequencies are less than $v_L^{min}$, and photon absorption will not occur.

Thus, in a 2 nm diameter monel NP, at "medical" frequencies, the two physical mechanisms, (a) and (b), mentioned in Sec. 3, will not work. The mechanisms (c) and (d) for absorption of THz radiation remain plausible.

An analysis of the distributions of the vibrational DOS of phonons in Cu-Ni alloys (with a broad peak in the vicinity of 5 THz, see Fig. 2*(a)* in Ref. 49) and dispersion curves of phonons in the 50at.%Cu50at.%Ni alloy (with a linear dependence up to a frequency of ≈5.3 THz, see Fig. 4*(b)* in Ref. 84) shows that in a 2 nm diameter monel NP, simultaneous absorption, by the Fermi electrons, of THz photons with frequencies of 0.35–0.55 THz and longitudinal phonons in the frequency range of ~4–5 THz is possible. Indeed, the wavelength $\lambda$ of longitudinal phonons with a frequency of $v$ ≈5 THz propagating along the diameter $D$ of the 2 nm diameter monel NP is $\lambda = v_L/v = 0.84$ nm; thus, $\lambda < D$, and such phonons can indeed be excited. Obviously, this confirms that the scheme of simultaneous absorption of photons with "medical" frequencies and longitudinal phonons with a frequency of ≈5 THz is possible in monel. The secondary phonon providing the relaxation of the excited electron would have a wavelength shorter than 0.84 nm.

Moreover, the relaxation of an excited electron is also possible due to electron-electron scattering, since the value of $l_{mfp}$ in monel is very small. One way or other, a dissipation of the energy acquired by the Fermi electron from a THz photon is provided.

Thus, the physical mechanisms (c) and (d) for absorption of THz radiation are realized in the 2 nm monel nanopartcile.

Summing up, in a 2 nm monel NP, in contrast to the 50at.%Au50at.%Pd alloy, a synergy of the properties (1), (2), and (3), mentioned above, in Subsec. 3.1.3, can be successfully realized.

At last, but not least, it should be noted that in the 2 nm monel NPs, their cooling is impossible due to spontaneous emission of THz photons, similar to that considered in Appendix A for GNPs. Indeed, in the 2 nm monel NPs, the value of $l_{mfp}$ is less than the NP's radius, therefore, the electrons excited by longitudinal phonons are relaxed in the NP's volume without emission of photons.

## 4 Prerequisites for the implementation of a THz-to-IR converter with a gelatin matrix

We will now calculate the parameters of a THz-to-IR converter with a gelatin matrix, a material which is technologically convenient and suggests a simple way to introduce NPs into the matrix material. Indeed, gelatin is a water-soluble protein derived from collagen. Commercially available NPs (in particular, GNPs) are supplied as a sol in water, or can be produced by laser ablation in water. Therefore, the introduction of NPs in the form of a sol in water into a water-soluble gelatin matrix and their uniform distribution in it should not be a problem.

However, technological simplicity brings with it a limitation: using gelatin, it is possible to operate at a frequency of 0.38 THz (wavelength $\lambda$: 789.5 μm), but not at higher frequencies, e.g., 4.2 THz ($\lambda$: 71.3 μm) and 8.4 THz ($\lambda$: 35.7 μm), since gelatin is opaque at these frequencies (see Fig. 9).

The transmission of gelatin in the wavelengths range ≈0.3–2.5 μm is quite large and overlaps with the 1.5–5 μm operating wavelengths range of commercial Mirage 640 P-series IR camera [31]. Gelatin transmission at a frequency of 0.38 THz ($\lambda$: 789.5 μm) is ≈75%, and at a wavelength of 1.5–2.5 μm it is ≈80–50%.

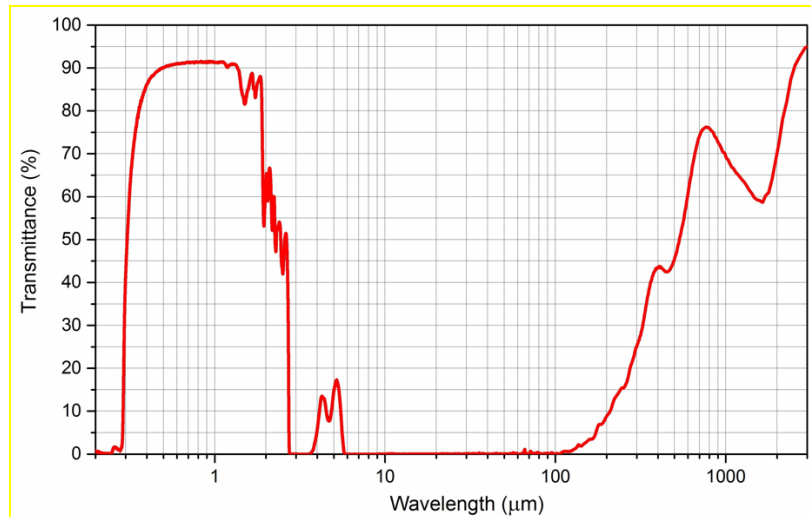

Fig. 9. Transmission spectrum of a gelatin film 180 μm thick.

Unlike gelatin, the Teflon®, polystyrene, polypropylene, and polyvinyl chloride matrices are transparent not only at 0.38 THz, but also at frequencies of 4.2 and 8.4 THz. (Refer to the transmission spectrum of polypropylene in Fig.4, for comparison with Fig. 9). Therefore, in addition to a contrast image of the skin cancer margins (at 0.38 THz), they are capable to produce a sharp image of cancer-affected skin area (at 4.2 and 8.4 THz).

Despite the abovementioned limitation inherent to gelatin, we estimate the parameters of this matrix in combination with 2 nm diameter NPs of monel. We hope that a THz-to-IR converter with such a matrix, coupled with an IR camera, could provide better resolution of the tumor tissue margin than commercially available THz imaging cameras do.

## 5 Calculations of the THz-to-IR converter parameters with a gelatin matrix

The calculations were carried out according to the method previously used in Refs. 1, 5.

### 5.1 *Power required to heat the NP to be seen by the IR camera; NP's heating and cooling times*

The problem was solved of temperature change in a metal NP placed inside a gelatin spherical shell as a result of heat releasing in the NP. The temperature change in the gold and monel spherical NPs was described by the heat conduction equation in spherical coordinates with the source function $q(r)$ taken into account:

$$\rho C \frac{\partial T}{\partial t} = \frac{1}{r^2} \frac{\partial}{\partial r}\left(\lambda r^2 \frac{\partial T}{\partial r}\right) + q(r), \qquad (7)$$

where $T(t, r)$ is the temperature, $\rho$ the volume density, $C$ the specific heat, $\lambda$ the thermal conductivity, $q$ the volume density of the heat source, $q = Q/[(4/3)\pi R_0^3]$ ($Q$ being the power at the heat source, $R_0$ the NP's radius), and $r$ the spherical radius. It was assumed that the thermal parameters did not depend on temperature and might be summarized as

$$0 \leq r \leq R_0: \quad \lambda = \lambda_1, \ \rho = \rho_1, \ C = C_1, \ q = \frac{Q}{(4/3)\pi R_0^3}, \qquad (8)$$
$$R_0 < r \leq R: \quad \lambda = \lambda_2, \ \rho = \rho_2, \ C = C_2, \ q = 0,$$

where $R_0$ is the NP's radius, and $R$ is the radius of the gelatin shell, $R \gg R_0$. In the course of numerical solution, $R = 500$ nm was taken as an effective "infinity", in view of much larger factual matrix thickness and the assumption that individual NPs are too distant to interfere.

The initial and edge conditions were as follows:

$$T(0, r) = T_R,$$
$$\left.\frac{\partial T(t, r)}{\partial r}\right|_{r=0} = 0, \ T(t, R) = T_R. \qquad (9)$$

A solution for Eqs. (7)–(9) was sought by the numerical method of lines [85] relative to $\Delta T = T - T_R$, where $T_R = 300$ K. To control the accuracy and time of calculations, we used the steady-state resolution of the problem:

$$\frac{1}{r^2} \frac{d}{dr}\left(\lambda r^2 \frac{dT}{dr}\right) + q(r) = 0, \ T'(0) = 0, \ T(R) = T_R,$$

which gave a relationship between power $Q$ and a maximum temperature rise $\Delta T_m$:

$$\Delta T_m = \frac{Q}{8\pi}\left(\frac{1}{R_0 \lambda_1} + \frac{2}{R_0 \lambda_2} - \frac{2}{R \lambda_2}\right).$$

On reaching the value of $\Delta T = 0.995\, \Delta T_m$, the heat source was turned off, and the problem was solved by cooling the particle down to 300 K. Numerical calculations were carried out with the material parameters given in Table 10.

Table 10. Parameters of materials used in numerical solutions of the heat conduction equation (7)

| Material | Volume density ($10^3$ kg·m$^{-3}$) | Specific heat (J·kg$^{-1}$·K$^{-1}$) | Thermal conductivity (W·m$^{-1}$·K$^{-1}$) |
|---|---|---|---|
| Nanoparticle: $D = 8.5$ nm gold sphere | $\rho_1 = 19.32$ | $C_1 = 133.7$ | $\lambda_1 = 70.5$ |
| Nanoparticle: $D = 2$ nm MONEL alloy 404 sphere | $\rho_1 = 8.91$ | $C_1 = 455.4$ | $\lambda_1 = 21.0$ |
| Matrix: gelatin | $\rho_2 = 1.32$ | $C_2 = 2039.9$ | $\lambda_2 = 0.21$ |

Data for gelatin (volume density, specific heat and thermal conductivity at 27°C) have been calculated using the formulas from Ref. 86.

For NPs, both their specific heats and thermal conductivities depend on the NP's size. According to Gafner et al. [87], the specific heat of GNPs is just slightly above that of bulk gold. We extrapolated the specific heat $C_1$ of the 8.5 nm diameter GNP from the data of Ref. 87 and estimated it to be 133.7 J·kg$^{-1}$·K$^{-1}$. The difference in the thermal conductivity is much more dramatic. In fact, the electron thermal conductivity on the nanoscale is much higher than the lattice thermal conductivity [88,89], however lower than that of the bulk metal. When the GNP's characteristic size (diameter) $D$ is smaller than the mean free path of electrons in the bulk gold, i.e., $l_{bulk} \approx 37.7$ nm at 300 K [77], the NP's thermal conductivity $\lambda_1$ can be estimated as follows [90,91]: $\lambda_1 \approx (\lambda_{1b}/l_{mfp}) \cdot D \approx 70.5$ W·m$^{-1}$·K$^{-1}$, where $\lambda_{1b} = 312.8$ W·m$^{-1}$·K$^{-1}$ is the thermal conductivity of bulk gold.

According to Ref. 87, in the temperature range of 150–800 K, the specific heat of nickel and copper face-centered cubic nanoclusters with diameter of up to 6 nm exceed the specific heat of these metals in bulk state, but by no more than 16–20%, even in the case of very small clusters. We assumed that the 2 nm diameter NPs from the nickel-copper alloy monel would also exhibit an increased specific heat. Therefore, in Table 10, we indicated a 10% increase in the specific heat of monel.

The source term in Eqs. (7) and (8), stationary in our assumption, accounts for the power $Q$ delivered to the particle. Obviously dependent on the power of the primary THz source, the relevant (threshold) values of $Q$ for our analysis are those which enable a heating, as a steady solution of Eq. (7) of the (embedded) GNPs and monel NPs to temperature levels detectable by the IR camera. We fix such reference threshold value at 12 mK, that is the reported temperature sensitivity of the Mirage 640 P-Series IR thermal imaging camera [31]. Individual GNPs and monel NPs, however, may largely vary in their ability to convert absorbed power into heat depending on the particle's emissivity factor $\alpha$, varying between 1 (ideal $Q$ to $\Delta T$ conversion) and almost zero. Without a clue as for the actual values of $\alpha$, we consider for reference purposes the values $\alpha = 1$ and $\alpha = 0.5$. The threshold $\Delta T$ values needed to yield visibility by the IR camera will be scaled by ×(1/$\alpha$), hence 12 and 24 mK, correspondingly, for the $\alpha$'s under discussion. The estimated powers $Q$, defined by the emissivity factors $\alpha$, versus the temperature rise $\Delta T$ for 8.5 nm diameter GNP and 2 nm diameter monel alloy spheres in gelatin spherical shells are given in Table 11. The heating and cooling times of the NPs are small enough to use a converter in the real time mode.

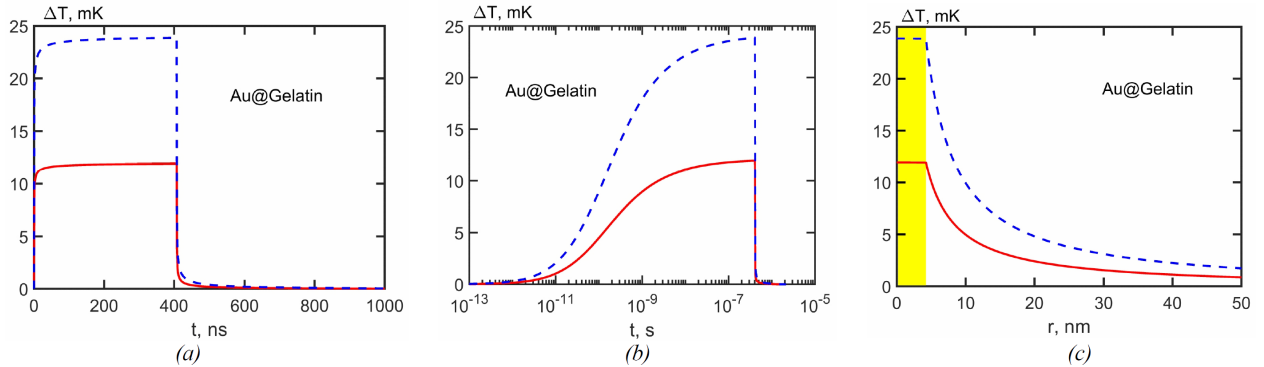

Fig. 10. *(a)* Temperature rise $\Delta T$ versus heating/cooling time $t$ for a GNP of diameter 8.5 nm in gelatin spherical shell for two values of $Q$ from Table 11. Upper curve corresponds to emissivity factor of GNP $\alpha = 0.5$, lower curves to $\alpha =1$. *(b)* Similar to the figure *(a)*, using the logarithmic time scale. *(c)* Radial distributions of excess temperatures throughout the GNP of diameter 8.5 nm (marked by a colour bar) and its embedding by gelatin spherical shell, for the power values of $Q$ from Table 11, corresponding to emissivity factor $\alpha=1$ (lower curve) and $\alpha=0.5$ (upper curve).

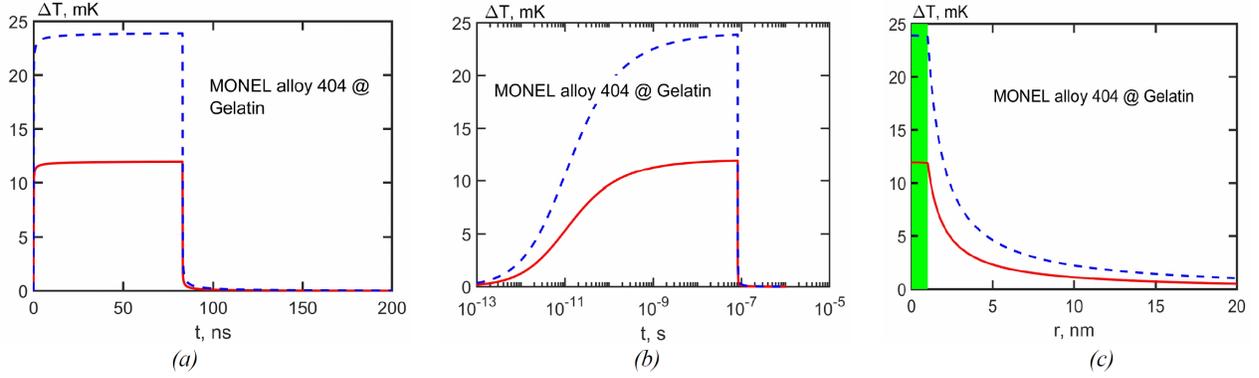

Fig. 11. The same as in Fig. 10 for a 2 nm diameter NP of monel alloy in gelatin spherical shell.

We calculated the temporal characteristics and spatial distributions of heat generated within the isolated NPs embedded in the matrix, as well as the power levels of the THz radiation required to maintain the GNPs in thermal equilibrium at, or above, the temperature sensitivity threshold of IR camera. Solving the heat conduction equation with parameters from Table 10 gave the following results:

Table 11. Calculated parameters of the gelatin matrix THz-to-IR converter

| Nanoparticle | Emissivity factor $\alpha$ of NP | $\Delta T$ (mK) | $Q$ (nW) | Heating time ($10^{-7}$ s) | Cooling time ($10^{-6}$ s) |
|---|---|---|---|---|---|
| $D$ = 8.5 nm gold sphere | 1 | 12 | 0.13 | 4.1 | 1.7 |
|  | 0.5 | 24 | 0.27 | 4.1 | 1.7 |
| $D$ = 2 nm MONEL alloy 404 sphere | 1 | 12 | 0.03 | 0.83 | 0.95 |
|  | 0.5 | 24 | 0.06 | 0.83 | 0.94 |

Figure 10 shows graphs of heating and cooling of a 8.5 nm diameter GNP in a gelatin matrix, and also a graph of the temperature distribution in and around the GNP. Figure 11 shows the same as in Fig. 10 for a 2 nm diameter monel NP.

### 5.2 *THz power threshold sensitivity for operating a THz-to-IR converter coupled with an IR camera*

On having discussed the "performances" of individual gold and monel NPs embedded in gelatin matrices for the THz-to-IR conversion, we turn to an assessment of a realistic device, to be composed of distributed GNPs or monel NPs and "viewed" by practically available IR cameras. Let us consider a spot on the converter matrix, which has to be mapped onto the pixel in the IR camera's focal plane array (FPA). As was mentioned in the Sec. 2, the IR camera operates in the close up mode, so that its objective's magnification nearly equals 1. Assuming that $d$ is the pixel's size, the spot's area is $d^2$.

We assume further on that the surface spot is expanded into a volume element with the same area $d^2$ and thickness $\delta$, inferior to the depth of focus $l_{\text{dof}}$ of the IR camera's objective. We specify that the volume element contains GNPs or monel NPs with the emissivity factor $\alpha$, heated by absorbing the THz energy. Those GNPs or monel NPs within $\delta \leq l_{dof}$ heated up to a temperature of 300 K + ($\Delta T_{bb}/\alpha$) will be perceived by the IR camera.

In the model considered, the minimal possible thickness of the film matrix is 1 μm, the diameter of the spherical shell covering the GNP or monel NP. For further assessments, gelatin film of 0.1 mm thickness would be an acceptable choice because this is less than the typical depth of focus $l_{dof}$ (~ 0.3 mm). Thin gelatin film could be applied on a thin substrate of any of considered plastic films.

In order to "sensibilize" each pixel, the number of NPs within the corresponding volume element must obviously be, at least, one. This ensures each pixel to be successfully addressable.

Assuming the number of NPs in the volume element equal to 1, the concentration of NPs in the matrix must be $N = 1/(d^2 \cdot \delta)$. Taking the Mirage 640 P-Series IR camera detector as an example, the pixel size is $d = 15$ μm, therefore, for the gelatin layer being 0.1 mm, the NPs concentration is $N \approx 4.44 \cdot 10^4$ mm$^{-3}$.

We can further on estimate the power of THz radiation source needed for a gelatin matrix of the 9.6 mm × 7.7 mm size and 0.1 mm thickness to work. The total amount of NPs in the matrix is $N \cdot 9.6$ mm·7.7 mm·0.1 mm $\approx 4.44 \cdot 10^4$ mm$^{-3}$·7.39 mm$^3 \approx 3.28 \cdot 10^5$. Typically for rough estimates one takes the emissivity factor $\alpha = 0.5$, then the power needed to heat such GNP by 24 mK is 0.27 nW, and the required power of the THz radiation source would be $3.28 \cdot 10^5 \cdot 0.27$ nW $\approx 88.6$ μW. Similarly, for a gelatin matrix with the monel NPs, the required power of the THz radiation source would be $3.28 \cdot 10^5 \cdot 0.06$ nW $\approx 19.7$ μW.

The values 88.6 μW and 19.7 μW can be considered as the conventional powers required for operating the THz-to-IR converter built around the gelatin matrix.

The power sensitivities of the THz-to-IR converters with gelatin and polystyrene matrices and monel NPs will be approximately the same as that of commercially available THz imaging cameras.

A few words about THz dosimetry. The frequency of 0.38 THz of interest to us is located near the common border (0.3 THz) of two established ICNIRP safety limits, [28] and [30], with sharply different limiting power densities of the incident radiation on the skin: 20 W·m$^{-2}$, average value over 6 minutes [28], and 1000 W·m$^{-2}$ with an allowable exposure duration of up to 30 ks, that is, up to ≈8.3 hours [30].

Based on the principle that observation/capturing should not introduce a change in the process being studied, and for the sake of patient safety, it is reasonable to choose the lower of the two power density values, 20 W·m$^{-2}$, but remember that this is an average value over 6 minutes. This means that if we choose the observation/capturing time less than 6 minutes, then during the observation/capturing we can increase the power density to 1000 W·m$^{-2}$ without harm to the patient's health, and then hold the interval until the next capturing – so that the average value over 6 minutes does not exceed 20 W·m$^{-2}$. Such a method would be useful for operating in the mode "Inspecting details of the object" (Table 6), since in this case the skin area to be examined (field of view), which is the "reflected radiation source" for the THz-to-IR converter, is quite small. In order to meet the operating power requirement of the THz-IR converter, it may be necessary for a short time to increase the irradiance above adopted 20 W·m$^{-2}$.

If a pulsed source of THz radiation is used, then the pulse duration should be slightly longer than the NP's heating time. By choosing the pulse frequency and the time intervals between periods of observation/capturing the skin area, it is possible not to exceed the established average value for 6 minutes, 20 W·m$^{-2}$.

The expected duration of the observation/capturing time can be several seconds (≤ 10 s).

Note that a power density of 1000 W·m$^{-2}$ is for the exposed skin area of 0.01 m$^2$ [30]. This corresponds to a 10 cm × 10 cm square of skin, a fairly large area. If typical exposed skin area is ~5 cm × 5 cm, then the limiting power density may be increased.

In laboratory practice, to study THz-induced effects in human skin, very high peak power densities were used: 47.2 MW·cm$^{-2}$ [92], 30 GW·cm$^{-2}$ [93-95], and 32 GW·cm$^{-2}$ [96]. For diagnostic purposes, the considered THz-to-IR converters could operate at significantly lower peak power densities, up to 1000 W·cm$^{-2}$. Such a power density will not harm the patient, as required by the ICNIRP Guidelines [30].

## 6 Discussion

We compared the parameters of THz-to-IR converters with different matrices and two types of embedded NPs: 8.5 nm diameter GNPs and 2 nm diameter copper-nickel MONEL alloy 404. The markedly different GNP parameters in the silicon matrix (see Table 1) attract attention: the power required to heat it to be seen by the IR camera is much larger than that in other matrices, very short heating/cooling times. This is because thermal conductivity of silicon is much higher than that of other matrix materials. We concluded that the option with a silicon matrix is unpromising and therefore did not consider the option with a monel NP in a silicon shell. However, for all other matrix materials, the calculated parameters for monel NPs are more profitable than for GNPs: thus, the heating times of monel NPs are ~5 times shorter, while the cooling times are ~2 times shorter. The power required to heat the monel NPs so that they can be seen by the IR camera are weaker by a factor of about 4, which promises to enhance the sensitivity of the THz-to-IR converter in terms of detectable THz power.

Calculated NP heating/cooling times show that the considered THz-to-IR converters would response to any object's evolution almost instantly, for all practical applications envisaged. Therefore, these converters could operate in real time. For technical operation, either a pulsed THz radiation source with pulse duration more than the NP heating time, with pauses between pulses longer than cooling time, or a continuous THz radiation source would be required. The conventional power required for operating the THz-to-IR converter built around the considered matrices are within accessible limits of contemporary THz radiation sources [97].

A gelatin matrix converter with commercially available 8.5 nm diameter GNPs is easier to manufacture, but it is only applicable for operation at the "medical" frequency of 0.38 THz (wavelength $\lambda$: 789.5 μm), at which gelatin is transparent. Unfortunately, gelatin is not transparent at practically important frequencies 4.2 THz ($\lambda$: 71.3 μm), 8.4 THz ($\lambda$: 35.7 μm), 5 THz ($\lambda$: 60 μm), and 10 THz ($\lambda$: 30 μm).

Due to confinement and uncertainty in electron energy levels, the absorption of metal NPs is broadband, which would allow to operate GNPs at frequencies of 4.2 and 8.4 THz, and to use monel NPs at 5 and 10 THz. This is advantageous when using these NPs embedded in plastic matrices that are transparent at the above operating frequencies of 4.2–10 THz, which makes it possible to provide both contrast and enhanced resolution of images of the same area of human skin. Naturally, when using plastic matrices with FLIR IR cameras, the power required to heat the NPs is larger than that necessary when using the Mirage IR camera (compare the data of Table 5 and Table 1).

Monel NPs could probably be produced in an aqueous medium by laser ablation of a piece of monel. The resulting monel NPs could easily be introduced into water-soluble gelatin, and a gelatin matrix with embedded nanoconverters of THz radiation to IR radiation could be obtained.

In Ref. 5, some of us outlined the idea of visualizing an image created in THz rays using NPs in a matrix. We set the magnification of the lens system of the IR camera equal to 1, and considered the case of one NP in the volume, the area of which corresponds to a IR camera detector pixel, and estimated the power that should be released in the NP so that the latter, being heated up, would overcome the temperature sensitivity threshold of the IR camera.

With this idea of visualizing, if we want to improve the image resolution, then, with a given magnification of the IR camera's lens system and for a given IR camera detector's temperature sensitivity, we must, first, reduce the pixel size of the IR camera detector; and, second, in order to adapt to the reduced pixel size, increase the concentration of NPs in the matrix. For example, the detector pixel size of the Mirage camera is $d_1 = 15$ μm; if the pixel size is reduced to $d_2 = 10$ μm, then the concentration of NPs in the matrix must be increased by $(d_1/d_2)^2 = 2.25$ times.

The development of the FPA detectors with improved characteristics is underway. New types of FPAs are now making their performance with temperature sensitivity 10 mK [98] and pixel's size 10 μm [99]. With this, the image resolution and power sensitivities of the THz-to-IR converters would be correspondingly enhanced.

In Sec. 1, we have already considered the factors that affect image quality: diffraction in the objective's lens, the pixel pitch size, and number of pixels in the IR camera detector. Now let's consider possible distortions in the THz-to-IR converter. Kuznetsov et al. [100-106] used the THz-to-IR converter with a topological pattern of split-ring resonators. Its drawbacks are the so-called "comet tail" and image blooming effects, which deteriorate the converter's response time and spatial resolution. No such effects are likely to occur in the THz-to-IR converters based on GNPs and monel NPs. For example, in the gelatin matrix, the edge of a reference cuboid accounted for one NP is equal to $(4.44 \cdot 10^4 \text{ mm}^{-3})^{-1/3} \approx 2.8 \cdot 10^4$ nm. The half of this distance, $1.4 \cdot 10^4$ nm, exceeds by far the maximum radii of radial distributions of temperature around heated GNPs shown in Fig. 10*(c)* (~50 nm) as well as of monel NPs shown in Fig. 11*(c)* (~20 nm). Hence, the temperature fields of neighboring heated either GNPs or monel NPs would not overlap. Moreover, the cooling times of GNPs and monel NPs are quite small. Therefore, no effects like the "comet tail" or the image blooming are a priori expected in the GNPs- and monel-NPs-based THz-to-IR converters.

We would also like to discuss some other approaches to the development of THz-to-IR converters. Bai et al. [107] demonstrated an ultrabroadband upconversion device, based on the integrated *p*-type GaAs/Al$_x$Ga$_{1-x}$As ratchet photodetector and GaAs double heterojunction LED. However, its photoresponse range, spanning from THz to NIR (4–200 THz), by far exceeded the frequencies suitable for biomedical research, e.g., 0.35–0.55 THz, capable to provide contrast margin of the cancerous tissue.

Alves et al. [108] developed a THz-to-IR converter focal plane array made of planar metamaterial structures of sensing elements. Upon THz absorption, the temperature of the sensing element increased, and the outward IR flux from the back side of the element was read by a commercial IR camera. Two structures were designed with sensing element sizes of 17.5 and 14.5 μm, minimum detectable temperature differences being about 40 mK, and thermal time constants were 170 ms and 100 ms for 3.8 THz and 4.75 THz operating frequencies, respectively. These parameters are also inferior to those of the THz-to-IR converters suggested in our work.

Fan et al. [109] proposed and demonstrated an approach which utilized all-dielectric metasurface absorbers consisting of an array of sub-wavelength silicon cylinders. Incident THz radiation was absorbed solely within the cylinders, converted to heat, and directly detected by an IR camera. The periodicity of the cylinder array was 330 μm, cylinder height 85 μm, radius 106 μm. The absorbance of cylinders was tuned to 0.6 THz, i.e., close to the frequencies suitable for biomedical research (0.35–0.55 THz); however, the cylinders' time constant was about 1 s, hence too slow for convenient practical applications. Due to this, and because the cylinders' radii being large, the temporal parameters and spatial resolution of this approach are inferior to those of nanoparticle-based THz-to-IR converters coupled with the IR cameras.

An advantage of the nanoparticle- and metamaterial-based THz-to-IR converters coupled with IR cameras is their simplicity; such devices will be able to operate without readout electronics nor cooling. The converters would be a simple attachment to commercial IR cameras permitting to convert the latter into THz cameras without any additional modifications. To offer a more complete overview, we would also mention an alternative design, in which a matrix with nanoparticles is mounted directly on a matrix with many microbolometers (as a development of the idea described in Ref. 110). An integration into a complementary metal oxide semiconductor (CMOS) technology along with its respective readout electronics and the use of Fresnel lenses instead of thick lenses for THz tumor imaging on the matrix with nanoparticles would reduce the dimensions of the system for imaging the human skin cancer.

The considered THz-to-IR converters could be used not only for in vivo studies, but also for in vitro research. In the transmission mode (for in vitro study of thin, clinically prepared tissue sample, see Fig. 12), the tissue sample areas containing cancer cells, rich in water, will strongly absorb the THz radiation, so that corresponding areas in the THz-to-IR converter's matrix will be in the THz "shadow" and thus come "cold" for the IR camera, on the background of image from healthy cells which let THz radiation pass and thus produce the "hot" image.

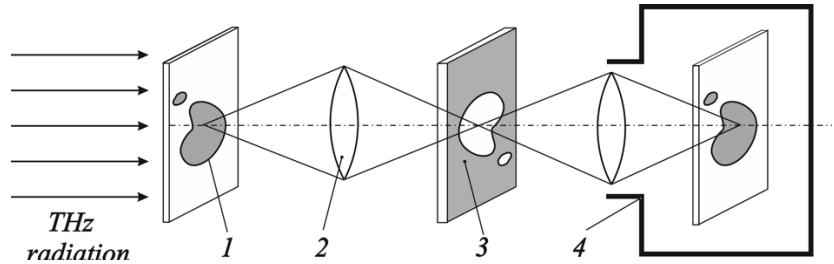

Fig. 12. Transmission mode for in vitro studies [4]: 1 – a tissue sample, 2 – THz objective with a magnification $M_1$, 3 – THz-to-IR converter, 4 – highly sensitive IR camera with magnification $M_2$. See [4] for details.

In contrast to the reflection geometry for in vivo imaging of oncopathology (see Fig. 2), which is focused on the study of the human skin cancer, the transmission mode studies can examine the tissue samples of different human organs.

We note that, in reference to Fig. 2, a simplified scheme with just a NIR laser and IR camera, without THz radiation nor THz-to-IR converter involved, may in principle serve for a preliminary tumor imaging. The disadvantage of this procedure, however, is that the margin of the tumor will be considerably blurred due to high thermal conductivity of the skin, because of the heat transfer from water in cancer cells to neighboring cells of normal biological tissue. The use of THz radiation reflected directly from cancer cells is expected to enable a more precise detection of the tumor margin.

## 7 Conclusion

Unlike ionizing X-rays, terahertz radiation is not ionizing and is therefore considered safe for living organisms. The THz medical imaging is expected to become a promising non-invasive technique for monitoring the human skin's status and for early detection of pathological conditions. In this context, the development of terahertz imaging setups emerged as one of the "hottest" areas in nanotechnology supported modalities for the cancer diagnostics. The THz-assisted diagnostics in reflection geometry allows for non-invasive (in vivo) THz imaging of skin cancer's surface features. It could be performed in situ, without time losses on standard in vitro histological tests, which require more time. In the reflection geometry, one could investigate skin only at the surface or to very small depth, to which the THz radiation, being strongly absorbed by water [111, 112], may penetrate. In the transmission geometry, one could study the clinically prepared tissue samples both of inner organs and of the skin.

Terahertz radiation can be useful for noninvasive monitoring of recovery from burns and wounds [113]. THz-to-IR converters have potential to be applied for visualization of healing process directly through clinical dressings and restorative ointments, minimizing the frequency of dressing changes. Experimental studies [113] show that in a reflective imaging configuration, the attenuation of EM radiation in wound dressings (2-pass propagation, a 1.5-mm layer of ointment is assumed) at frequencies of 0.25–0.6 THz is much smaller than that at infrared frequencies. This means that at a frequency of 0.38 THz, which is favorable for the visualization of skin oncopathologies, the use of the THz-to-IR converters for wound monitoring could also be successful.

In addition, the THz imaging in reflection geometry has perspectives to be used in dermatology, e.g. for monitoring the treatment of psoriasis, since this allows to avoid the direct contact with the skin, contrary e.g. to a case of ultrasound investigation.

It is also very important from the end user's point of view that the THz imaging approach is highly cost effective compared to the magnetic resonance method.

Finally, according to the review articles [114–118], relatively few studies on the biological effects of THz radiation had been performed in 7.0–10.0 THz frequency range. Since the 8.4 THz and 10.0 THz operating frequencies of the THz-to-IR converters we propose fall into this poorly studied frequency region, the converters we have considered would contribute to replenishing information on biological effects in this THz frequency region.

## Disclosures

All the authors declare that they did not have neither financial nor material support for their research related with this work, through which they might have a potential conflict of interest.
This article is based on a modified and expanded version of the SPIE Proceedings manuscript "Terahertz-to-infrared converter based on the polyvinylchloride matrix with embedded gold nanoparticles" [9].

## Acknowledgements

The authors thank Prof. O. P. Cherkasova, Dr. N. A. Usov and Dr. D. S. Sitnikov for valuable discussions, and Dr. O. V. Dymar for helpful advice on assessing gelatin parameters. This work was partially supported by funding from the Academy of Finland (grant projects 314639, 351068 and 325097), ATTRACT II META-HiLight project funded by the European Union's Horizon 2020 research and innovative programme under grant agreement No.101004462, the Leverhulme Trust and The Royal Society (Ref. no.: APX111232 APEX Awards 2021).

## Appendix A: Spontaneous emission of THz photons by gold nanoparticles

Spontaneous emission of THz photons by GNPs with diameter less than ≈8 nm is a manifestation of the size effect and quantization of energy levels of vibrational modes (longitudinal phonons) under confinement conditions. Consider a compression wave (longitudinal phonon with momentum $q^*_{vm}$) propagating along the NP's diameter $D$ (Fig. 13(a)). Longitudinal phonons with energy $E_{vm}$ and momentum $q^*_{vm}$ are permanently absorbed by Fermi electrons (assumed to be adequately treated within the free-electron model) with momentum $p_F = 1.27 \cdot 10^{-19}$ g·cm·s$^{-1}$ and energy $E_F = 5.53$ eV [63]. The excited (upon absorption) electrons move along the direction characterized by the momentum $s$, the modulus of which is $s = [2m \cdot (E_F + E_{vm})]^{1/2}$, and an angle $\gamma$ to the phonon momentum (see Fig. 13(a)): $\gamma = \arccos\,[(2m \cdot E_{vm} + q^{*2}_{vm})/(2s \cdot q^*_{vm})]$. For the longitudinal phonon dominant in gold (with energy $E_{vm} \approx 17.4$ meV, and magnitude of the momentum $q^*_{vm} \approx 1.13 \cdot 10^{-19}$ g·cm·s$^{-1}$), the angle $\gamma$ is 63.6°.

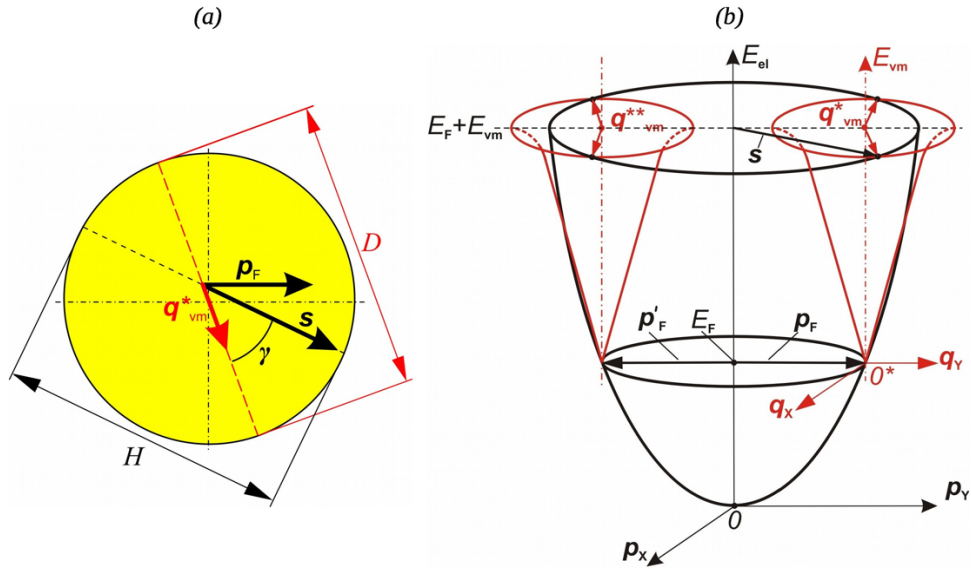

Fig. 13. Absorption of the vibrational mode (longitudinal phonon) with momentum $q^*_{vm}$ by the Fermi electron with momentum $p_F$ in GNP. The excited electron with momentum $s$ moves along the chord $H$.

Excited electrons, due to the Coulomb interaction with positive ions of gold, could induce secondary longitudinal phonons with momentum collinear to $s$; this would be the relaxation path of excited electrons. However, in NPs smaller than ≈8 nm, this mechanism is impossible, and relaxation occurs due to the emission of a photon.

The first reason contributing to emission is the difference in the quantization steps in momenta and energies for longitudinal phonons propagating along the NP's diameter $D$ and along the chord $H$, collinear to $s$. The momentum steps for phonons propagating along the diameter and along the chord are, respectively, $h/D$ and $h/H$ ($h$ being the Planck's constant). Since $D > H$, the momenta steps are different; consequently, the energy steps for these two directions of phonon propagation do not coincide either. Due to a mismatch of the quantized energy levels, the transfer of energy from the excited electron to the secondary phonon is impossible (the Fermi electron received energy from the primary phonon quantized with a certain step, and it cannot generate a secondary phonon, in which the energy would be quantized with a different step).

Another reason precluding an excitation of secondary phonon and hence favoring the THz emission is the following. As the NP's diameter decreases, the gap between the energy levels of longitudinal phonons for the direction along the vector $s$ increases, eventually becoming so large that the gap exceeds the $FWHM_L$ (Figure 14 shows the threshold stage of this situation, see details below). As a result, the energy level of an excited electron "hangs" between the levels of longitudinal phonons for the direction $s$: there are none to which it could pass the energy. Consequently, the electron relaxes upon scattering at the NP boundary, emitting a photon (the electron cannot leave the NP, since its energy is inferior to the work function of gold, ≈4.8 eV).

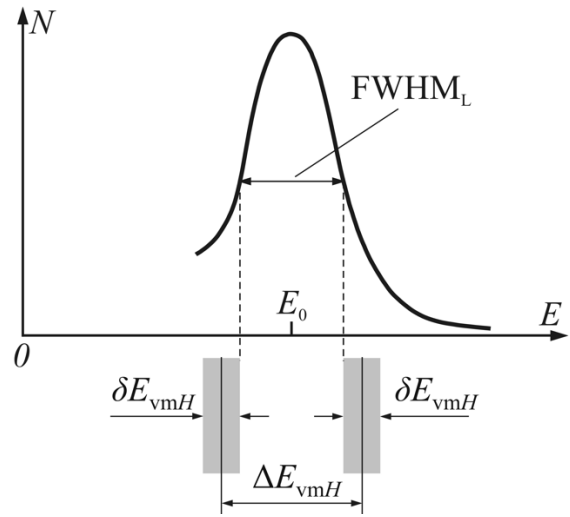

Fig. 14. The threshold stage in which the energy step for vibrational modes $\Delta E_{vmH}$ propagating along the chord $H$ with momentum $s$ exceeds the $FWHM_L$ of the energy distribution of longitudinal phonons in GNP.

For the direction "along the chord $H$", the steps in momentum and energy of the vibrational modes are, respectively, $h/H$ and $\Delta E_{vmH} \approx v^*_L \cdot (h/H)$, where $v^*_L$ is the speed of sound for the energy range of longitudinal phonons within the $FWHM_L$. $v^*_L$ is less than the nominal speed of sound in gold, $v_L = 3.23 \cdot 10^5$ cm·s$^{-1}$, due to the bending down of the dispersion curve of longitudinal phonons near the Brillouin zone boundary, where the phonon energies corresponding to $FWHM_L$ are distributed.

Table 12. Correlation of the catalytic activity of a GNP and its diameter $D$.

| $D$ (nm) | $\Delta E_D$ (meV) | $\delta E_{vmH}$ (meV) | $FWHM_L/\Delta E_D$ | Note |
|---|---|---|---|---|
| 1.2 | 3.45 | 0.55 | 0.81 | Highest catalytic activity |
| 1.5 | 2.76 | 0.44 | 1.01 | Decrease in catalytic activity with increasing diameter D |
| 2.5 | 1.655 | 0.26 | 1.69 | |
| 5.0 | 0.83 | 0.13 | 3.37 | |
| 8.0 | 0.52 | 0.08 | 5.38 | |
| 8.5 | 0.49 | 0.08 | 5.71 | Lack of catalytic activity |
| 10.0 | 0.41 | 0.065 | 6.83 | |

Obviously, the mismatch of energy levels is somehow tempered by the Heisenberg uncertainty relation. We take as a criterion of "total mismatch" between the energy levels of phonons to be potentially excited the situation at which the two consecutive energy levels definitely move apart by more than the $FWHM_L$ in the longitudinal phonon DOS within the tolerance criteria imposed by the uncertainty relation. Such threshold situation is depicted in Fig. 14, where $\delta E_{vmH}$ is calculated using the Heisenberg's uncertainty relation for the phonon's momentum and the coordinate, $\delta E_{vmH} \geq v^*_L \cdot h/(2\pi H)$. The inequality that determines the corresponding threshold chord length (so that at smaller values the spontaneous emission of photons begins to occur) reads as follows:

$$\Delta E_{vmH} > FWHM_L + (\delta E_{vmH}/2) + (\delta E_{vmH}/2) = FWHM_L + \delta E_{vmH} \geq FWHM_L + v^*_L \cdot h/(2\pi H). \quad (10)$$

Substituting the value $\Delta E_{vmH} \approx v^*_L \cdot (h/H)$ into the left side of inequality (10), we arrive at:

$$v^*_L \cdot (h/H) > FWHM_L + v^*_L \cdot (h/2\pi \cdot H) \text{ or } H < [1 - (1/2\pi)] \cdot v^*_L \cdot (h/ FWHM_L). \quad (11)$$

Should this inequality be satisfied, a significant fraction of excited electrons won't be able to relax with the generation of secondary phonons – they will emit photons, and the emission intensity will be at maximum.

Substituting the numerical values of the quantities ($v^*_L \approx 10^5$ cm·s$^{-1}$; $FWHM_L \approx 2.8$ meV), we obtain an estimate of the chord length providing the maximum intensity of spontaneous emission of photons: $H < 1.2$ nm. Since $H < D$, the estimate for the diameter $D \leq 1.2$ nm would be a valid condition to yield yet an appreciable spontaneous emission of photons, in a situation when the vibration mode levels are separated (by $\Delta E_{vmH}$, cf. Fig. 14) quite sparsely. With an increase in the diameter of a GNP from 1.2 to 8 nm, the emission intensity will decrease and stop at $D \approx 8$ nm.

It is namely by the spontaneous emission of THz photons that we explain the nature of the catalytic activity observed in GNPs (Refs. 119–121) with a diameter less than $\approx 8$ nm (see Ref. 57 and Table 12). The idea is that the emission of THz photons promote the catalysis of chemical reactions on the surface of GNPs. The emission of photons will be accompanied by cooling of the NPs insofar as they are not thermally insulated.

The cessation of photon emission at $D \approx 8$ nm could also be a manifestation of the energy dissipation of an excited electron due to its multiple scattering by other electrons, if the NP's size becomes comparable to the average electron mean free path in the NP. Judging by the experimental data on the catalytic activity of GNPs depending on their diameter [119–121], the average electron mean free path in GNPs is not less than $\approx 8$ nm.

The higher the multiplicity of laying the energy step of the longitudinal phonons $\Delta E_D$ within the $FWHM_L$ width, that is, the larger the ($FWHM_L/\Delta E_D$) ratio, the higher the probability of superposition of the energy level expanded to $\delta E_{vmH}$ on the level of the primary phonon, which will lead to electron relaxation with the generation of a secondary phonon (with energy of the primary phonon). And the lower the intensity of spontaneous emission of photons, the lower the NP's catalytic activity and the lower the degree of its cooling.

The data of Refs. [119–121] suggest that in GNPs in the range of diameters $D < 8$ nm, the inequality $l_{mfp} < D$ is not satisfied (otherwise the excited electron would scatter and emission would be impossible). The works [122, 123] have shown (in silver, a chemical analogue of gold) that even in sub-10-nm silver particles, the crystalline lattice is preserved, and its distortion occurs only on the surface of the particles. This is a situation when $l_{mfp} = D$, and $D < l_{bulk}$ (in bulk gold, $l_{bulk}$ is quite large, $l_{bulk}$ = 37.7 nm [77]). All this suggests that in the GNPs tested in Refs. [119–121], electron scattering occurs only at the NP's boundaries (an electron "flies" the NP from boundary to boundary without scattering in its core), i.e., the conditions for photon emission are satisfied. But this also strengthens the assumption that, in GNPs, the cessation of emission with an increase in the GNP's size is due precisely to the entry of the levels of vibrational modes for the direction along the chord – into the $FWHM_L$ wide energy range.

In connection with our hypothesis of spontaneous THz photon emission and cooling of GNPs smaller than 8 nm, we note that when studying the size effect of heating GNPs for RF hyperthermia (at ~13.56 MHz), [57], we did not find experimental data on the successful use of GNPs smaller than 5 nm. This may be an indirect confirmation of the hypothesis: it was not possible to effectively heat a GNP smaller than 5 nm because, simultaneously with heating by RF radiation, it was also cooled due to the spontaneous emission of THz photons, which carried away energy from the GNP. This correlates with the data that catalysis starts at GNPs smaller than ~8 nm.

According to our model estimates [57], in GNPs smaller than 5 nm, there may still be a heating mechanism due to longitudinal phonons propagating over the surface of GNPs, but its role decreases with decreasing GNP diameter. Apparently, it is overcome by cooling due to the spontaneous emission of THz photons, the intensity of which increases with a decrease in the GNP diameter, and is accompanied by an increase in the catalytic activity of the GNP.

## Appendix B: Features of confinement of electrons and phonons in GNPs

Metal NPs are quantum objects: in them the energy and momenta levels of electrons and phonons are quantized. Another property of quantum objects, subordination to the Heisenberg uncertainties relations, has to be taken into account when considering the absorption of EM photons by NPs.

With practically interesting GNP sizes, the "competition" of discreteness and uncertainties of energies $\delta E_{el}$ and momenta $\delta p_{el}$ of electrons is "wined" by uncertainties: they exceed the quantization steps $\Delta E_{el}$ and $\Delta p_{el}$, i.e., the discreteness of energies and momenta can be ignored. For example, in the 8.5 nm diameter GNPs considered, the uncertainties of levels of electron energies $\delta E_{el}$ and momenta $\delta p_{el}$ exceed the steps in their energies $\Delta E_{el}$ and momenta $\Delta p_{el}$ by more than 2 orders of magnitude (see estimates in Appendix C).

Since the manifestations of confinement in NPs are rarely considered in papers, we allow ourselves to discuss them in some detail. First, consider a $D$ diameter GNP at a temperature 0 K. Figures 15, *a)*, *b)* and *c)* show the positions $E'_F$ and $E''_F$ of the Fermi level at $T = 0$ K, corresponding to the minimal uncertainty $\delta p_{el} = h/(2\pi D)$ in the electron momentum at the Fermi level. What is important: inside the $\delta E_{el}$ zone, the position of the Fermi level is not determined. It can be anywhere within this zone.

In a 8.5 nm diameter GNP, a gap between the electron energy levels $\Delta E_{el} \approx 0.39$ meV is much smaller than the uncertainty in the electron energy $\delta E_{el} \approx 108.5$ meV. Thus, the energy uncertainty zones of several (not just neighboring) electron levels overlap. This would mean that several electrons with the same spin, stemming from different energy levels, could be in the same state in what concerns energy. Is there a conflict between the Pauli principle and the Heisenberg uncertainty relation here?

We have considered and qualitatively explained the observed and possible interactions of photons with electrons and phonons, using representations within the framework of the free electron model. Therefore, it is not surprising that in a small volume of metal NP, they lead to a conflict between the Pauli principle and the consequences of the Heisenberg uncertainty relations.

We could speculatively assume that the movement of electrons in small NPs is correlated in such a way that electrons with the same spin and energy do not come close to each other, being screened by other electrons. But it is clear that for a more rigorous description of the interaction of photons, electrons, and phonons under confinement conditions, one should proceed to theoretical models related to the concepts of many-body effects. We restrict ourselves to the possible interactions considered here qualitatively, assuming that this is sufficient for an engineering problem.

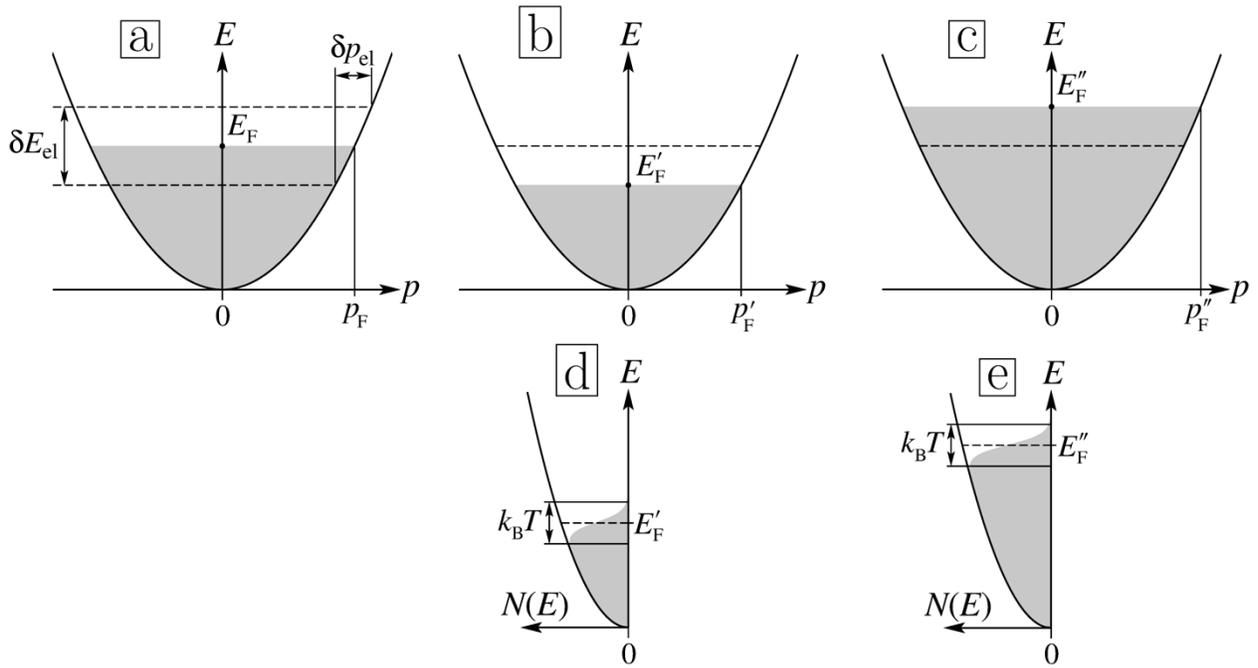

Fig. 15. Uncertainty in the position of the Fermi level $E_F$ in a GNP as a manifestation of confinement (*a, b, c*; $T = 0$ K) and the distribution of the DOS of electrons (*d, e*; $T > 0$ K) due to thermal smearing of the Fermi level. The areas marked in gray are the areas of energy occupied by electrons.

Let now the NP's temperature $T > 0$ K. Figures 15, *d)* and *e)* show the distributions of the DOS of electrons corresponding to the cases of Figs. 15, *b)* and *c)*, respectively. Figures 15, *b)* and *d)* are for the case when $E'_F = E_F - (\delta E_{el}/2)$; whereas the Figs. 15, *c)* and *e)* are for the case when $E''_F = E_F + (\delta E_{el}/2)$. Each of the positions $E'_F$ and $E''_F$ of the Fermi level has thermal smearing $\sim k_B T$. The electron levels within $\sim k_B T$ are incompletely filled.

Let us estimate the number of electron levels in a 8.5 nm diameter GNP in the band of incompletely occupied states, $\sim k_B T/2$ wide below the Fermi level. At $T = 300$ K, $k_B T/2 \approx 13$ meV, the gap between the electron levels is $\Delta E_{el} = 0.39$ meV; hence, within the band $\sim k_B T/2$, there are 13 meV/0.39 meV = 33 underoccupied levels, where electrons could end up after scattering. On the other hand, electrons in this band could absorb energy and momentum and be excited. So the photon energy will dissipate – turn into heat, and heat the GNP.

## Appendix C: Reference summary of data

Here we provide a reference summary of data on the 8.5 nm diameter GNPs, THz radiation suitable for imaging human skin cancer, and confinement-induced parameters of electrons and longitudinal phonons in GNPs (Table 13).

Let's make some useful evaluations. According to estimates for the 8.5 nm diameter GNPs, the uncertainty $\delta E_{el} \approx 108.5$ meV. In these GNPs, the uncertainties of discrete levels of electron energies $\delta E_{el}$ and momenta $\delta p_{el}$ exceed the steps in their energies $\Delta E_{el}$ and momenta $\Delta p_{el}$ by more than 2 orders of magnitude.

In the same GNP, the minimum uncertainties in the energies and momenta of longitudinal phonons, $\delta E_{vm}$ and $\delta q_{vm}$, are $2\pi$ times smaller than the gaps $\Delta E_{vm}$ and $\Delta q_{vm}$, respectively.

Such a difference in the ratio of quantization steps and uncertainties for electrons and phonons is a manifestation of a large difference in the speed of Fermi electrons and the speed of sound in gold: they differ by 3 orders of magnitude.

Let us estimate the uncertainty in the energy of longitudinal phonons $\delta E_{vm}$ corresponding to the frequencies $\nu = n_{vm} \cdot (v_L/D)$, see Eq. (3). We get a "smearing" of phonon levels, on which an excited electron could relax. The uncertainty in the energy of longitudinal phonons in the same 8.5 nm diameter GNP, – in the linear section of the dispersion curve in gold, is equal to: $\delta E_{vm} \approx v_L \cdot \delta q_{vm}$, where $\delta q_{vm}$ is the uncertainty in the momentum of a longitudinal phonon propagating along the GNP's diameter $D$.

The uncertainty $\delta q_{vm}$ is estimated from the Heisenberg relation: $\delta q_{vm} \cdot D \geq (h/2\pi)$, whence: $\delta q_{vm} \geq h/(2\pi \cdot D) = 1.2 \cdot 10^{-21}$ g·cm·s$^{-1}$. Then the uncertainty in the energy of the longitudinal phonon will be no less than: $\delta E_{vm} \approx v_L \cdot \delta q_{vm} = 0.25$ meV.

A gap between energy levels of longitudinal phonons is estimated as $\Delta E_{vm} \approx v_L \cdot \Delta q_{vm} = v_L \cdot (h/D) \approx 1.57$ meV. For comparison, the uncertainty in the phonon energy is $\delta E_{vm} \approx v_L \cdot \delta q_{vm} \geq v_L \cdot h/(2\pi \cdot D) = \Delta E_{vm}/2\pi$. This means that the minimum uncertainty in the phonon energy $\delta E_{vm}$ is $2\pi$ times smaller than the gap $\Delta E_{vm}$.

The quantization step for momentum of longitudinal phonons is $\Delta q_{vm} = h/D = 7.8 \cdot 10^{-21}$ g·cm·s$^{-1}$. For comparison, the uncertainty in the momentum of a longitudinal phonon is $\delta q_{vm} \geq h/(2\pi \cdot D) = \Delta q_{vm}/2\pi = 1.2 \cdot 10^{-21}$ g·cm·s$^{-1}$. This means that the minimum uncertainty in the phonon momentum $\delta q_{vm}$ is $2\pi$ times smaller than the gap $\Delta q_{vm}$.

Using the Kubo formula (Refs. 61,62; see also Appendix 1 in Ref. 57), one can estimate the gap $\Delta E_{el}$ between the electron energy levels in a 8.5 nm diameter GNP: $\Delta E_{el} \approx (4/3) \cdot (E_F/N)$. Here, $E_F = 5.53$ eV is the Fermi energy of gold [63]. At $D = 8.5$ nm, $N \approx 1.89 \cdot 10^4$, the gap $\Delta E_{el} \approx 0.39$ meV. For comparison, the uncertainty of the electron energy levels is $\delta E_{el} \approx \delta p_{el} \cdot v_F = 108.5$ meV. It is more than 2 orders of magnitude greater than the gap $\Delta E_{el}$ between the electron energy levels.

The electron momentum step $\Delta p_{el} \approx \Delta E_{el}/v_F = 4.5 \cdot 10^{-24}$ g cm·s$^{-1}$. Hence, the estimated above uncertainty in the electron momentum $\delta p_{el} \geq 1.2 \cdot 10^{-21}$ g·cm·s$^{-1}$ is more than 2 orders of magnitude larger than the electron momentum step $\Delta p_{el}$.

Table 13. A 8.5 nm diameter GNP data and estimates of its confinement-induced parameters

| $a_{Au}$ (nm) | $E_F$ (eV) | $v_F$ (cm/s) | $v_L$ (cm/s) | $D$ (nm) | $N$ | $\nu$ (THz) | $E_{ph}$ (meV) | $p_{ph}$ (g·cm/s) | $d$ (nm) | $FWHM_L$ (meV) |
|---|---|---|---|---|---|---|---|---|---|---|
| 0.408 | 5.53 | $1.4·10^8$ | $3.23·10^5$ | 8.5 | $1.89·10^4$ | 0.38 | 1.57 | $8.4·10^{-26}$ | 121.0 | ≈2.8 |

| $l_{bulk}$ (nm) | $\Delta p_F$ (g·cm/s) | $\Delta p_{el}$ (g·cm/s) | $\delta p_{el}$ (g·cm/s) | $\Delta E_{el}$ (meV) | $\delta E_{el}$ (meV) | $\delta\nu$ (THz) | $\Delta q_{vm}$ (g·cm/s) | $\delta q_{vm}$ (g·cm/s) | $\Delta E_{vm}$ (meV) | $\delta E_{vm}$ (meV) |
|---|---|---|---|---|---|---|---|---|---|---|
| 37.7 | $1.8·10^{-23}$ | $4.5·10^{-24}$ | $1.2·10^{-21}$ | 0.39 | 108.5 | 26.3 | $7.8·10^{-21}$ | $1.2·10^{-21}$ | 1.57 | 0.25 |

Top row: GNP data and THz radiation characteristics:
$a_{Au}$ – lattice constant of gold;
$E_F$ – Fermi energy in gold;
$v_F$ – Fermi electron velocity in gold;
$v_L$ – the velocity of propagation of longitudinal phonons in gold;
$D$ – diameter of the GNP;
$N$ – number of gold atoms in a 8.5 nm diameter GNP;
$\nu$ – the THz radiation frequency, acceptable to identify tumor margins;
$E_{ph}$ – photon energy of THz radiation;
$p_{ph}$ – the THz radiation photon momentum;
$d$ – the skin depth in gold at the frequency $\nu$;
$FWHM_L$ – full width at half maximum of a peak in the density of longitudinal phonon frequency distribution in gold.

Bottom row: confinement-induced parameters of electrons and longitudinal phonons in a GNP:
$l_{mfp}$ – the mean free path of electrons in gold at 300 K;
$\Delta p_F$ – an increase of the Fermi electron's momentum in the course of absorbing a THz photon;
$\Delta p_{el}$ – momentum step of electrons in a $D$ diameter GNP;
$\delta p_{el}$ – uncertainty in the momentum of a Fermi electron due to confinement;
$\Delta E_{el}$ – the gap between the energy levels of electrons in a $D$ diameter GNP;
$\delta E_{el}$ – uncertainty in the Fermi energy and electronic levels due to confinement;
$\delta\nu$ – uncertainty in the Fermi level and electronic levels in units of frequency;
$\Delta q_{vm}$ – momentum step of longitudinal phonons propagating along the GNP's diameter $D$;
$\delta q_{vm}$ – uncertainty in the momentum of a longitudinal phonon propagating along the GNP's diameter $D$;
$\Delta E_{vm}$ – the gap between the energy levels of longitudinal phonons propagating along the GNP's diameter $D$;
$\delta E_{vm}$ – uncertainty in the energy of the longitudinal phonon level due to the uncertainty in the momentum of the longitudinal phonon $\delta q_{vm}$.